\documentclass[11pt,nofootinbib,tightenlines,preprintnumbers,amsmath,amssymb,longbibliography,notitlepage]{revtex4-1}
\usepackage{epsfig}
\usepackage{graphicx}
\usepackage{latexsym}
\usepackage{booktabs}
\usepackage{graphicx}
\usepackage{color}
\usepackage{footnote}
\usepackage{hyperref}

\definecolor{darkred}{rgb}{0.55, 0.0, 0.0}
\definecolor{crimsonglory}{rgb}{0.75, 0.0, 0.2}

\newcommand{\be}{\begin{equation}}
\newcommand{\ee}{\end{equation}}
\newcommand{\ba}{\begin{eqnarray}}
\newcommand{\ea}{\end{eqnarray}}
\newcommand{\bi}{\begin{itemize}}
\newcommand{\ei}{\end{itemize}}

\newcommand{\<}{\langle}
\renewcommand{\>}{\rangle}

\newcommand{\la}{\label}

\newcommand{\ten}[1]{\times 10^{#1}}
\newcommand{\tabref}[1]{Table~\ref{#1}}
\renewcommand{\eqref}[1]{Eq.~(\ref{#1})}

\begin{document}
\title{{The pion quasiparticle in the low-temperature phase of QCD}}

\author{Bastian B. Brandt$^1$, Anthony Francis$^2$,  Harvey B. Meyer$^{3,4}$ and Daniel Robaina$^3$\vspace{0.2cm}}

\affiliation{
$^1$Institut f\"ur theoretische Physik, Universit\"at Regensburg, D-93040 Regensburg, Germany\\
$^2$Department of Physics \& Astronomy, York University, 4700 Keele St, Toronto, ON M3J 1P3, Canada\\
$^3$PRISMA Cluster of Excellence \& Institut f\"ur Kernphysik,
Johannes Gutenberg-Universit\"at Mainz, D-55099 Mainz, Germany\\
$^4$ Helmholtz~Institut~Mainz, Johannes Gutenberg-Universit\"at Mainz, D-55099 Mainz, Germany 
\vspace{0.2cm}
}

\preprint{MITP/15-042}

\date{\today}

\begin{abstract}
We investigate the properties of the pion quasiparticle in the
low-temperature phase of two-flavor QCD on the lattice with support
from chiral effective theory.  We find that
the pion quasiparticle mass is significantly reduced compared to its
value in the vacuum, by contrast with the static screening mass, which
increases with temperature.  By a simple argument, near the chiral
limit the two masses are expected to determine the quasiparticle
dispersion relation.  Analyzing two-point functions of the axial
charge density at non-vanishing spatial momentum, we find that the
predicted dispersion relation and the residue of the pion pole are
simultaneously consistent with the lattice data at low momentum.  The
test, based on fits to the correlation functions, is confirmed by a
second analysis using the Backus-Gilbert method.
\end{abstract}
\maketitle

\section{Introduction\la{sec:intro}}

Identifying the spectrum of excitations of strongly interacting matter
at finite temperature is of central importance to understanding the
nature of the medium. These excitations are encoded as poles in
thermal correlation functions.  For some quantum numbers, the
existence of weakly interacting probes coupling to local operators
make it possible to measure the properties of these excitations
experimentally, at least in principle. The prime example of such a
probe is the photon.  In practice, a medium is created in heavy-ion
collisions which appears to reach thermal equilibrium locally, the
temperature decreasing with time. Therefore a weighted average of
thermal photon or dilepton spectra over the spacetime history of the
`fireball' is obtained (see
e.g.\ \cite{Bratkovskaya:2014mva,Ruan:2014kia} and Refs.\ therein).

In the low-temperature phase, it is natural to ask how close the
properties of the excitations are to those of the known hadrons at
zero temperature. Viewed globally, the spectrum does not appear to
change much until temperatures close to the transition temperature are
reached, where the rapid crossover to a deconfined and chirally
symmetric phase takes place. This conclusion is based on the success
of the hadron resonance gas model in describing equilibrium properties
of the medium (particularly the equation of state and quark number
susceptibilities) computed in lattice
QCD~\cite{Bazavov:2012jq,Borsanyi:2013bia,Borsanyi:2011sw}, and on its
success in describing particle yields in heavy-ion
collisions~\cite{BraunMunzinger:2011ta,Stachel:2013zma}.
However, reliable information about individual excitations is sparse.

Here we extend our study~\cite{Brandt:2014qqa} of the pion at finite
temperature in two-flavor lattice QCD with support from a thermal
chiral effective theory~\cite{Pisarski:1996mt,Son:2001ff,Son:2002ci}.
In \cite{Brandt:2014qqa} we performed a temperature-scan on
$16\times 32^3$ ensembles; here we focus on one temperature
($T=170\,{\rm MeV}$) on a fine lattice ($24\times 64^3$) with high
statistics, for which we also have a reference zero-temperature
ensemble at the same bare parameters.  Our study shows that
the zero-temperature pion mass `splits' at finite temperature into a
lower pion quasiparticle mass and a higher pion screening mass,
 \ba
&  T=0: & \qquad\qquad\qquad\qquad\qquad\qquad \textrm{ pion mass} = 267(2)\,{\rm MeV} 
 \nonumber\\
 \!\!\!\!\!\!\! \!\!\!\!\!\!\! \!\!\!\!\!\!\! 
& & \qquad \qquad \qquad \qquad \qquad \qquad \qquad \swarrow ~~~~~~~~~~~~ {\searrow} 
\nonumber \\
& T=169{\rm MeV}:& \qquad \textrm{quasiparticle mass} = 223(4){\rm MeV} 
  \qquad  \quad 
 \textrm{screening mass} =303(4){\rm MeV}.
 \nonumber
 \ea
While the quasiparticle mass is the real-part of a pole of the
retarded correlator $G_R(\omega,|{\bf p}|=0)$ of the pseudoscalar
density in the frequency variable, the (static) screening mass is a
pole of $G_R(\omega=0,{\bf p})$ in the spatial momentum $|{\bf p}|$
and represents an inverse spatial correlation length.  The pion
quasiparticle mass can be extracted model-independently near the
chiral limit due to the dominance of its contribution to the two-point
function of the axial charge density and of the pseudoscalar density.
By contrast with the mass, we find that the decay constant associated
with the pion quasiparticle practically retains its zero-temperature
value.

By a simple argument, the dispersion relation of the pion at low
momenta is given by a single parameter $u$ (see
\eqref{eq:pion_disprel} below), which in the chiral limit corresponds
to the group velocity of the excitation.  As
in~\cite{Brandt:2014qqa}, we determine this parameter as the ratio of
the quasiparticle mass to the screening mass, $u\approx 0.74$.  As a
new aspect, we then test whether the so-determined parameter $u$
correctly predicts the momentum-dependence of the pion energy by
looking at the two-point function of the axial charge density at
non-vanishing spatial momentum. An important observation is that the
chiral Ward identities also predict the residue of the pion pole in
the axial-current two-point functions. Due to the difficulty of
extracting real-time information from Euclidean correlation functions,
testing simultaneously the predictions for the pole and the residue
proves to be essential to improve the discriminative power of the
analysis. Since the chiral predictions are only expected to be valid
at sufficiently small momenta, we also provide an estimate of the
range of validity of the effective theory.

Our analysis method of lattice correlation functions is based on fits,
where the ansatz is motivated by the chiral effective theory at small
frequencies and on perturbation theory at high frequency. We also
present an alternative analysis, which starts by generating
model-independently a locally averaged spectral function by following
the Backus-Gilbert inversion method~\cite{Backus1,Backus2,Press:2007zz,
  doi:10.1080/01630568708816267,schomburg1987convergence,kirsch1988backus}. In
a second step, a pion pole contribution with the predicted dispersion
relation is assumed, allowing us to obtain an estimate for the
residue. The advantage of this alternative analysis is that we do not
have to formulate an explicit ansatz for the spectral density of the
non-pion contributions. This point is particularly relevant since at
finite spatial momentum, axial-vector excitations do contribute to the
two-point function of the axial charge density.

The paper is organized as follows.  Section \ref{sec:theo} contains an
overview of the theory expectations concerning the two-point functions
of the axial current at finite temperature. Sections~\ref{sec:latsetup} and~\ref{sec:latcal} present the lattice QCD
calculation, and our conclusions are given in section \ref{sec:concl}.
In appendix \ref{sec:tensor}, the tensor structure of the
axial-current two-point functions at finite temperature is given; in
appendix \ref{sec:apdx_residue}, we derive the contribution of the
pion to the four independent tensor structures, thus determining all
the relevant residues. Finally, appendix~\ref{sec:tables} contains a table with the 
lattice correlator data.

\section{Theory background\la{sec:theo}}

We work in the Euclidean path integral formalism, and
our notation and conventions follow those used in  \cite{Brandt:2014qqa}.
The vector and axial-vector vectors and the pseudoscalar density are given by 
\be\la{eq:currdef}
V^a_\mu(x) = \bar{\psi}(x)\gamma_\mu \frac{\tau^a}{2}\psi(x), \qquad A^a_\mu(x)
 = \bar{\psi}(x)\gamma_\mu \gamma_5 \frac{\tau^a}{2}\psi(x),\qquad P^a(x) 
 = \bar{\psi}(x)\gamma_5 \frac{\tau^a}{2}\psi(x),
\ee
where $\psi$ is the isospin-doublet quark field.
In  appendix \ref{sec:tensor}, we provide a decomposition in momentum space of the
Lorentz structure of the two-point functions of the axial current. For
a general momentum $p$, they are entirely described by four `form factors', which 
in the rest frame of the thermal medium are functions of $p_0$ and ${\bf p}^2$.  
At zero-temperature, the four functions reduce to two functions of $p^2$, one longitudinal and
one transverse.  The partially-conserved axial current (PCAC) relation relates the two-point function
$\<P^a(x)P^b(0)\>$ of the pseudoscalar density, as well as the
$\<A^a_\mu(x)P^b(0)\>$ correlation functions to the aforementioned form factors.

In this work, we investigate the following static screening correlators,
\begin{eqnarray}\la{eq:GAs}
\delta^{ab}\,G^\text{s}_{A} (x_3, T) &=& \int dx_0\; d^2x_{\perp} \left<A^{a}_3(x) A^{b}_3(0)\right>, \\
\delta^{ab}\,G^\text{s}_{P} (x_3, T) &=& \int dx_0\; d^2x_{\perp} \left<P^a(x)P^b(0)\right>,
\la{eq:GPs}
\end{eqnarray}
where $x_{\perp} = (x_1,x_2)$. Time-dependent correlators with a general spatial momentum ${\bf p}$ will also play a crucial role,
\begin{eqnarray}
\la{eq:GA}
\delta^{ab}\,G_A (x_0, T, {\bf p}) &=& \int d^3x~ e^{i {\bf p\cdot x}}\left<A^{a}_0(x)A^{b}_0(0)\right>, \\
\delta^{ab}\,G_P (x_0, T, {\bf p}) &=& \int d^3x~ e^{i {\bf p\cdot x}} \left<P^a(x)P^b(0)\right>.
\la{eq:GP}
\end{eqnarray}
They are related by Fourier transformations to the form factors defined in appendix \ref{sec:tensor}, 
for instance
\be
\la{eq:GAsFF}
G^\text{s}_{A} (x_3, T) = \int \frac{dp_3}{2\pi}\, e^{-ip_3x_3}\, \Pi^{\rm L,l}(0,p_3^2).
\ee
The correlators $G_A (x_0, T, {\bf 0})$ 
and $G^\text{s}_{A} (x_3, T)$ are only sensitive to the longitudinal form factor $\Pi^{\rm L,l}$;
these were the cases considered in \cite{Brandt:2014qqa}. At non-vanishing momentum however, 
the correlator $G_A (x_0, T, {\bf p})$ is sensitive to three independent 
form factors $\Pi^{\rm T,l}$, $\Pi^{\rm M}$ and  $\Pi^{\rm L,l}$.

\begin{figure}[t]
\begin{center}
\includegraphics[width=.55\textwidth]{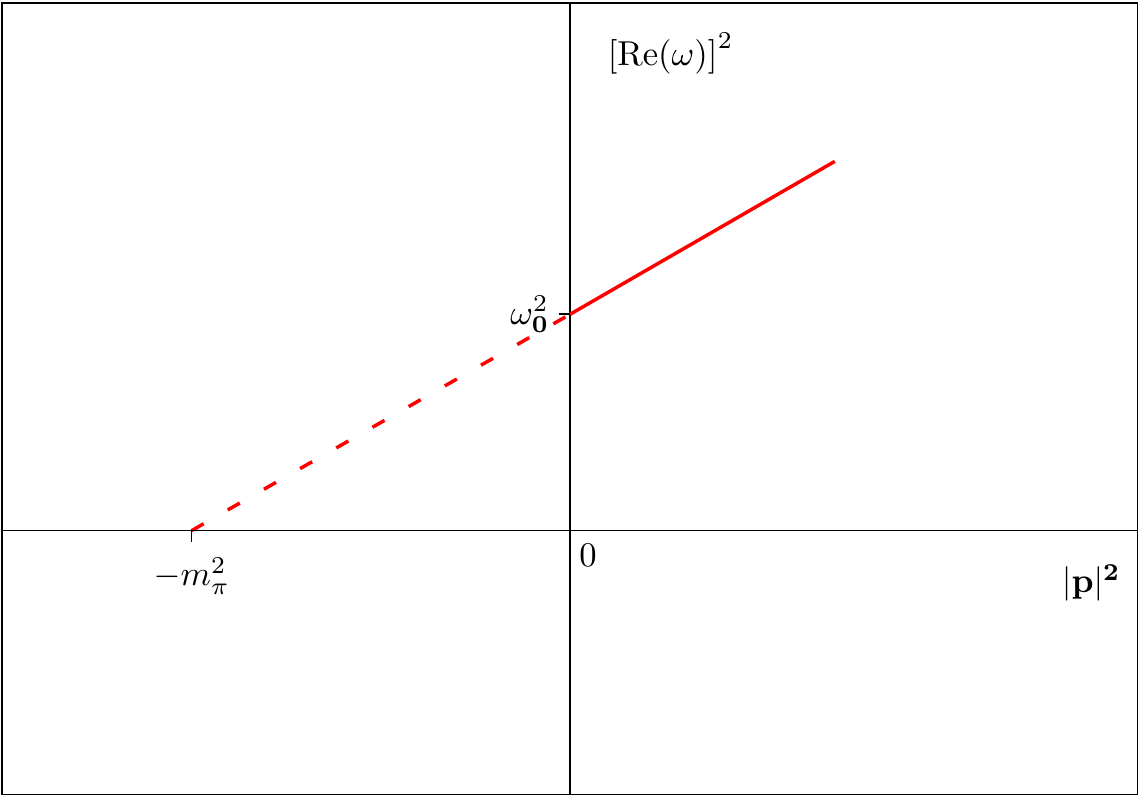}
\end{center}
\caption{Trajectory in $(\omega, {\bf p})$-plane 
of the pion pole in the pseudoscalar retarded correlator. 
At negative ${\bf p}^2$ the pole corresponds to the pion screening mass, 
at positive ${\bf p}^2$ it corresponds
to the pion quasiparticle. The slope is the value of $u^2(T)$.}
\label{fig:residue_picture}
\end{figure}

At long distances, the screening correlator $G^\text{s}_{A} (x_3, T)$ is given by 
\be\la{eq:Gsasy}
G^\text{s}_{A} (x_3, T) \stackrel{|x_3|\to\infty}{= }  \frac{1}{2}f_\pi^2 m_\pi \,e^{-m_\pi|x_3|},
\ee
which defines the screening pion mass $m_\pi$ and the associated decay constant\footnote{The 
normalization convention is such that at zero temperature $f_\pi\approx 92\,{\rm MeV}$.} $f_\pi$.
The Gell-Mann--Oakes--Renner relation 
\be\la{eq:GOR}
f_\pi^2 m_\pi^2 = - m\<\bar\psi\psi\>
\ee
holds to leading order in the chiral expansion.
From Eqs.\ (\ref{eq:GAsFF}) and (\ref{eq:Gsasy}), 
the low-momentum analytic structure of the longitudinal form factor $\Pi^{\rm L,l}$ reads
\be
\Pi^{\rm L,l}(0,{\bf p}^2)
= \frac{f_\pi^2 m_\pi^2}{{\bf p}^2 +m_\pi^2},\qquad \quad {\bf p}\to 0.
\ee
More generally, expanding the denominator in the frequency, 
\be
{\bf p}^2 +m_\pi^2 ~~\longrightarrow~~ {\bf p}^2 +m_\pi^2 + \frac{1}{u^2}\omega_n^2+\dots,
\ee
it follows that
a quasiparticle (pole in the retarded correlator as a function of frequency) with the dispersion relation~\cite{Son:2001ff,Son:2002ci}
\be
\omega_{\bf p} = u(T) \sqrt{m^2_\pi + {\bf p}^2} 
\label{eq:pion_disprel}
\ee
exists at low momenta\footnote{In this argument, the imaginary part of the frequency-pole is neglected. 
A more sophisticated argument is required to show that the damping rate of the pion quasiparticle
is indeed parametrically subleading~\cite{Son:2002ci}.}. The remarkable aspect is that the parameter $u$ determines
both the (real part of the) dispersion relation of the quasiparticle
and the ratio of the quasiparticle mass to the screening mass. 
A graphical interpretation of the dual role of the parameter $u$ is
given in Fig.~\ref{fig:residue_picture}.  Here the trajectory in the
frequency-momentum plane of a pole in the retarded correlator of the pseudoscalar density\footnote{
Recall 
that the momentum-space Euclidean correlator $G_E(\omega_n,{\bf p})$ 
is related to the retarded correlator via
  $G_R(i\omega_n,{\bf p}) = G_E(\omega_n,{\bf p})$ for $\omega_n\geq 0$~\cite{Meyer:2011gj}.}
corresponds to a static screening state at ${\bf p}^2 = -m_\pi^2$, and
to a real-time quasiparticle at small positive ${\bf p}^2$. 
In \cite{Brandt:2014qqa}, the parameter $u$ was determined 
using lattice correlation functions at vanishing spatial momentum via the two estimators
\begin{eqnarray}\la{eq:um}
u_m &=& \left[-\frac{4m^2_q}{m^2_\pi} \left.\frac{G_P(x_0,T,0)}{G_A(x_0,T,0)}\right|_{x_0 = \beta/2}\right]^{1/2} \\ 
u_f  &=& \frac{f^2_\pi m_\pi}{2 G_A(\beta/2, T, 0)\sinh(u_f m_\pi \beta/2)}.
\la{eq:uf}
\end{eqnarray}
In doing so, the parametric dominance of the pion in the
time-dependent Euclidean correlator at small quark masses is
exploited. Good agreement was found between $u_f$ and $u_m$ at
$T\simeq 150\,{\rm MeV}$ for a zero-temperature pion mass of
$305\,{\rm MeV}$.  Any departure of $u$ from unity clearly represents
a breaking of Lorentz invariance due to thermal effects.  In this
work, one of our goals is to test whether the parameter $u$ determined from the ratio of the quasiparticle to the
screening mass,  as in~\cite{Meyer:2011gj}, really does predict the dispersion relation of the
quasiparticle, as in \eqref{eq:pion_disprel}.
In order to carry out this goal, we  perform an analysis of the time-dependent 
Euclidean correlator $G_A(x_0, T, {\bf p})$ in terms of the spectral function $\rho^A$. They are related as follows,
\be
G_A(x_0, T, {\bf p}) = \int^\infty_0 d\omega~ \rho^{A}(\omega, {\bf p}) \frac{\cosh(\omega(\beta/2-x_0))}{\sinh(\omega \beta/2)}.
\label{eq:spectral_def}
\ee

First we recall that at zero temperature,
the Lorentz structure of the axial-current two-point function  
implies the following momentum dependence of the pion pole contribution,
\be
G_A(x_0, T=0, {\bf p}) 
\stackrel{|x_0|\to\infty}{\sim} \text{Res}(\omega^0_{\bf p})\frac{ e^{-\omega^0_{\bf p} |x_0|}}{2\omega^0_{\bf p}}
\ee
where the residue is here given by
\be
 \text{Res}(\omega^0_{\bf p}) = (f^0_{\pi} \,\omega^0_{\bf p})^2,
\qquad \quad \omega^0_{\bf p}  = ({\bf p}^2 + (\omega^0_{\bf 0})^2)^{1/2}.
\label{eq:res_T0}
\ee
In terms of the spectral functions, this correlator corresponds to 
\be
\rho^{A}(\omega, T=0, {\bf p}) = \text{Res}(\omega_{\bf p}^0)\, \delta(\omega^2 - (\omega^0_{\bf p})^2) + \dots
\label{eq:spec_T0}
\ee
At the other end of the spectrum, in the high frequency region, a leading-order perturbative calculation 
(see for instance \cite{Aarts:2005hg}) yields
\be
\rho^{A}(\omega, T, {\bf p}) = \theta(\omega^2 - 4m^2 - {\bf p}^2) \frac{N_c}{24 \pi^2} ({\bf p}^2 + 6m^2), \qquad \omega\to\infty.
\label{eq:asymptotic}
\ee
We note that at non-zero momentum, the correlator $G_A(x_0, T=0,{\bf p})$ 
receives contributions from axial vector mesons via the transverse form factor (see appendix \ref{sec:tensor}),
\be
G_A(x_0, T=0, {\bf p}) = \int \frac{dp_0}{2\pi} e^{-ip_0x_0} \Big[\frac{{\bf p}^2}{p_0^2+{\bf p}^2}\Pi^{\rm T}(p_0^2+{\bf p}^2) 
+ \frac{{p_0}^2}{p_0^2+{\bf p}^2}\Pi^{\rm L}(p_0^2+{\bf p}^2)\Big].
\ee
The spectral functions associated with the form factors $\Pi^{\rm T}$
and $\Pi^{\rm L}$ are measured experimentally in $\tau$
decays~\cite{Davier:2005xq}.  Since $\Pi^{\rm T}$ describes by itself
the two-point function of spatial components of the axial current at
vanishing spatial momentum, it cannot contain the pion pole. The
latter is entirely contained in the form factor $\Pi^{\rm L}$.

At finite temperature, the pion pole appears in all three form factors contributing to $G_A(x_0, T, {\bf p})$.
Altogether, the pion contribution to the spectral function $\rho^A$ is predicted to have the form 
$\rho^A(\omega,T,{\bf p}) = {\rm Res}(\omega_{\bf p}) \delta(\omega^2-\omega_{\bf p}^2)$, 
with the dispersion relation given by
\eqref{eq:pion_disprel} and  the residue by (see appendix \ref{sec:apdx_residue})
\begin{eqnarray}\la{eq:res}
\text{Res}(\omega_{\bf p}) &=& f^2_\pi (m^2_\pi + {\bf p}^2).
\end{eqnarray}
For later use we also define the pion quasiparticle decay constant $f_\pi^t$ via
\be
\text{Res}(\omega_{\bf 0}) = (f_\pi^t \,\omega_{\bf 0})^2.
\ee
The contribution to the Euclidean correlator then reads
\be
G_A(x_0, T, {\bf p})
= \frac{\text{Res}(\omega_{\bf p})}{2\omega_{\bf p}} \frac{\cosh(\omega_{\bf p}(\beta/2-x_0))}{\sinh(\omega_{\bf p} \beta/2)}+\dots.
\label{eq:residue}
\ee
Whether the residue determined through fits to lattice correlation functions agrees
with \eqref{eq:res} provides a cross-check that the low-energy effective description is working.

\section{Lattice setup\label{sec:latsetup}}
In this section we describe the analysis performed on a finite
temperature ensemble of size $24 \times 64^3$ with two degenerate
dynamical light flavors. The short direction is interpreted as time
and therefore the temperature is $T=1/24a = 169(3)$MeV while the
spatial extent amounts to $L=64a=3.1\,{\rm fm}$. 
The fields admit thermal boundary conditions in time and periodic 
boundary conditions in space.
We use the Wilson plaquette action \cite{Wilson:1974sk}
and the O(a) improved Wilson fermion action with a non-perturbatively
determined $c_\text{sw}$ coefficient \cite{Jansen:1998mx}. The
configurations were generated using the MP-HMC algorithm
\cite{Hasenbusch:2001ne} following the implementation described in
\cite{Marinkovic:2010eg} based on L\"usher's DD-HMC package
\cite{CLScode}. In addition, we use a $128 \times 64^3$, effectively
zero temperature ensemble that was made available to us through the
CLS effort (labelled as O7 in \cite{Fritzsch:2012wq}) with
all bare parameters identical to our
finite temperature ensemble. The pion mass takes a value of
$m_\pi=270$MeV \cite{Fritzsch:2012wq} such that $m_\pi L=4.2$. This
additional zero temperature test ensemble allows us to compare thermal
observables in a straightforward manner with their corresponding
``effective zero-temperature" value calculated in the O7 ensemble.

\subsection{PCAC mass}

In order to check that our thermal ensemble indeed yields the same
physical quark mass as its corresponding zero-temperature counterpart (O7), we
use the definition of the quark mass based on the PCAC
(partially conserved axial current) relation
\cite{Bochicchio:1985xa,Luscher:1996ug}
\begin{equation}\la{eq:PCAC}
m_{\rm PCAC}(x_3) = \frac{1}{2} \frac{\int dx_0 d^2x_{\perp} 
\left<\partial^{\rm imp}_3 A^{\rm imp}_3(x) P(0)\right>}{\int dx_0 d^2x_{\perp} 
\left<P(x)P(0)\right>} \qquad x_\perp = (x_1, x_2)
\end{equation}
where in the improvement process
\begin{equation}
A^a_{\mu} \to A^{\text{imp},a}_{\mu} = A^a_{\mu} + ac_A \partial^\text{imp}_\mu P^a.
\label{eq:imp}
\end{equation}
The derivative $\partial^{\rm imp}_\mu$ is the improved lattice
discretized version of the derivative following
\cite{Guagnelli:2000jw}. The non-perturbatively calculated coefficient
$c_A$ was taken from \cite{DellaMorte:2005se}. Note that since
relation (\ref{eq:PCAC}) stems from an operator identity, we are free
to choose the direction of measurement. On the thermal ensemble, the
spatial directions are longer; therefore, by measuring along these
directions we obtain a longer plateau and thus more accurate
determinations of the PCAC mass. The extraction can be carried out by
performing a fit to a constant in the range where a plateau is
observed. Its central value and error, given in Table \ref{table:O7param},
are in very good agreement with the ones quoted in \cite{Engel:2014eea}.

\begin{figure}[h!]
\begin{center}
\includegraphics[width=.7\textwidth]{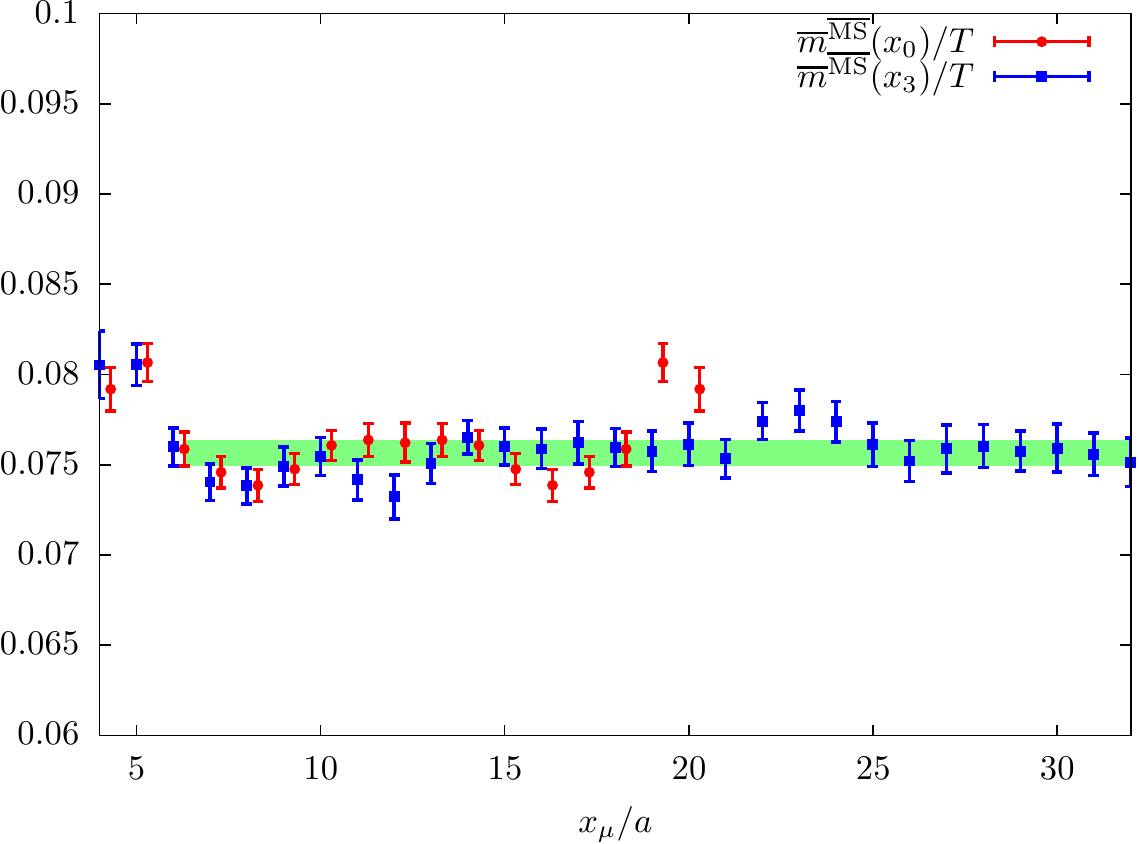}
\end{center}
\caption{The PCAC masses. The renormalization factors $Z_A$ and $Z_P$
  are included, as well as the conversion factor from the SF to
  $\overline{\text{MS}}$ at a scale of $\mu=2\text{GeV}$, which amounts to
  0.968(20) \cite{Fritzsch:2012wq}. We also  plot the result
  along the $x_0$-direction to show that indeed both are
  compatible. This can be interpreted as a check that cutoff effects
  are indeed small for this value of the lattice spacing.}
\end{figure}

\begin{table}[h!]
\centering
\begin{tabular}{l r}
\hline \hline
$6/g^2_0$ & 5.50\\
$\kappa$ & 0.13671 \\
$c_{\rm sw}$ & 1.751496 \\ \hline
$T_{{N_\tau}=24}$ [MeV] & 169(3) \\
$T_{{N_\tau}=128}$ [MeV] & 32(1) \\
$a$ [fm]~~\cite{Fritzsch:2012wq} & 0.0486(4)(5) \\
$Z_A$ ~~\cite{Fritzsch:2012wq} & 0.793(4) \\
$Z_P$ ~~\cite{Fritzsch:2012wq} & 0.5184(53) \\
$\overline{m}^{\overline {\rm MS}}/T (\mu=2{\rm GeV})$ & 0.0757(7) \\ \hline\hline
\end{tabular}
\caption{Summary of the main parameters for the $24 \times 64^3$ finite temperature ensemble
 as well as for the $128 \times 64^3$ zero temperature ensemble labelled as O7 in \cite{Fritzsch:2012wq}. 
The quark mass is computed at and normalized with the $T = 1/24a$ temperature.
The statistics collected for two-point functions is respectively 360 and 149 configurations at $N_\tau=24$ and 
$N_\tau=128$, with respectively 64 and 16 point sources per configuration, exploiting the translational invariance 
of the system.}
\label{table:O7param}
\end{table}

\subsection{Pseudoscalar and axial-vector correlators}

Our goal is to calculate the temperature-dependent coefficient $u(T)$ that parametrizes
the pion dispersion relation (\ref{eq:pion_disprel}).
In \cite{Brandt:2014qqa}, we defined two estimators $u_f$ and $u_m$
that yielded consistent results up to $T\simeq 170$MeV for the case of two
degenerate light flavors with
$\overline{m}^{\overline{\text{MS}}}(\mu=2\text{GeV}) \sim 15$ MeV;
at that quark mass, the crossover region is located around $T_C \simeq 211$ MeV
\cite{Brandt:2013mba}. In the thermal ensemble we are analyzing here,
we have $\overline{m}^{\overline{\text{MS}}}(\mu=2\text{GeV}) =
12.8(1)$MeV (see \tabref{table:O7param}), and therefore expect a
slightly lower value of the transition temperature. Nevertheless, this
should not affect the applicability of the chiral expansion around
$(T, m=0)$ with $T<T_C$, as discussed in \cite{Brandt:2014qqa}.

We use the correlators defined in Eqs. (\ref{eq:GAs}--\ref{eq:GP})
with the spatial momenta given by 
\be
{\bf p}= {\bf p}_n \equiv (0,0,2\pi n/L).
\ee
The improvement of the axial current was already introduced in
\eqref{eq:imp}. Note that since all two-point functions belong to
the adjoint (or \emph{isovector}) representation of SU($N_f$)
$(N_f=2)$, the contributions of quark disconnected diagrams cancel
out. The renormalization program is carried out such that
\begin{eqnarray}
G_A (x_0, T, {\bf p}) &=& (Z_A(g^2_0))^2 G_A(x_0,g^2_0, T, {\bf p})\\
G_P (x_0, T, {\bf p}) &=& (Z_P(g^2_0))^2 G_P(x_0,g^2_0, T, {\bf p})
\end{eqnarray}
and similarly for the screening correlators;
the value of the coefficients $Z_A$ and $Z_P$ can be found in \tabref{table:O7param}.

\section{Analysis of lattice two-point functions\la{sec:latcal}}

\begin{figure}[t!]
\begin{center}
\includegraphics[width=0.68\textwidth]{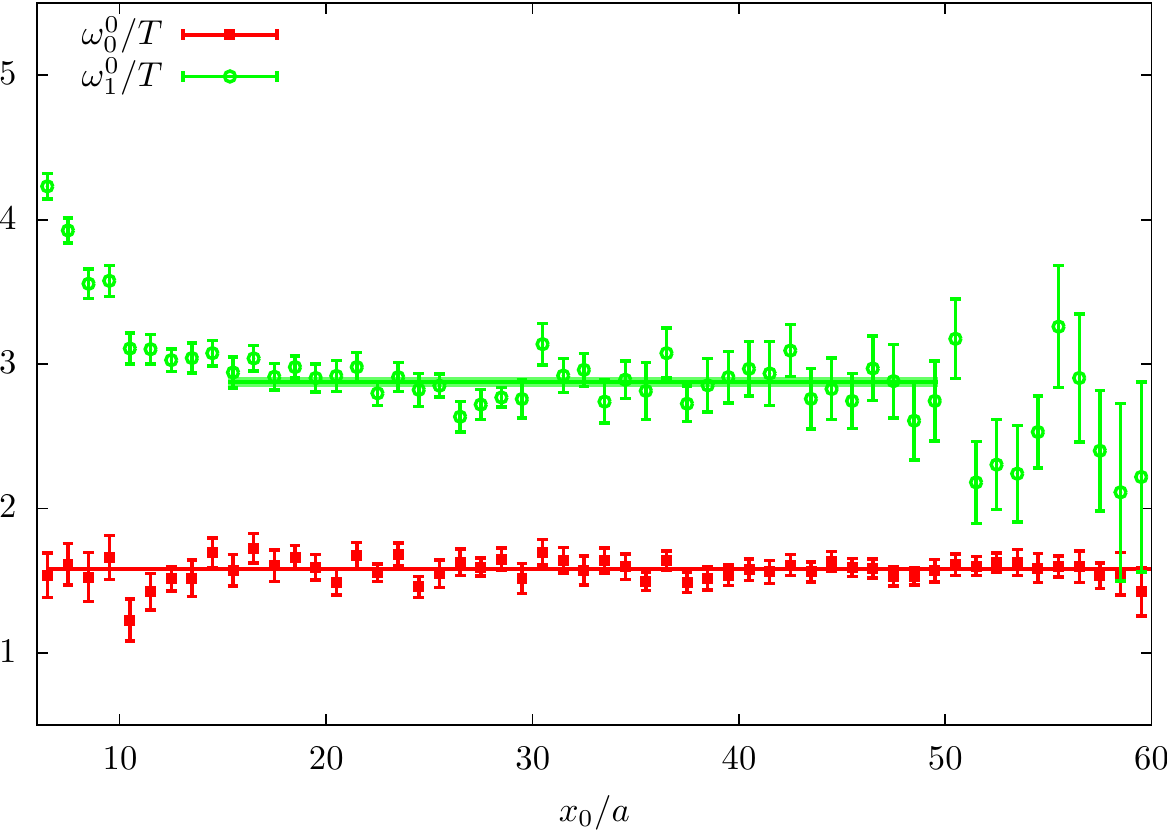} 
\end{center}
\caption{Effective `cosh' masses for the O7 zero-temperature ensemble in the
  $\left<A_0A_0\right>$ channel for $n=0,1$. 
The values of $\omega^0_n\equiv \omega^0_{{\bf p}_n}$ are given in units of the temperature  $T=1/24a$
corresponding to our thermal ensemble.}
\label{fig:u_T0}
\end{figure}

After the preliminary work presented in the previous section, we turn
to the analysis of correlation functions in order to extract the pion properties.

\subsection{The zero-temperature case}
As a benchmark we analyze zero-temperature data on the O7 ensemble.
Here we are able to obtain the pion energy $\omega^0_{\bf p}$ by fitting a constant to the
effective mass. The pion energy corresponding to ${\bf p} = 0$ and
${\bf p}=(0,0,2\pi/L)$ can be read off from the  plot in
Fig.~\ref{fig:u_T0}. The dominance of the pion contribution,
particularly in the zero-momentum case, is clearly very
strong. Performing a linear fit to $(\omega^0_{\bf p})^2$ as a function of ${\bf p}^2$, 
we obtain for the slope $u^2(T\simeq0) = 1.01(6)$, consistently with Lorentz invariance.
The decay constant ${f^0_\pi}$, defined by \eqref{eq:res_T0}, 
indeed turns out to be independent of the momentum.

Once the ground state dominates the correlator $G_A(x_0, T\simeq0,
{\bf p})$, one-state cosh fits of the form $A_1
\cosh(\omega^0_{{\bf p}}(T/2-x_0))$ with $T=128a$ are applied and the
results are summarized in \tabref{tab:cosh_T0}. The values for
$\omega_{\bf 0}^0$ and ${f^0_\pi}$ are in very good agreement with the ones
quoted in \cite{Engel:2014eea}.

\begin{table}[h!]
\centering
\begin{tabular}{c|c|c|c|c|c}  
$n$ & $A_1/T^3$ & $\omega^0_{{\bf p}_n}/T$ & $\chi^2/d.o.f$ & $f_{\pi}^0/T$ & $\text{Res}(\omega_{\bf p})$ \\ \hline 
0 & $8.4(3)\ten{-3}$ & $1.579(12)$  & $0.05$ & $0.599(8)$ & 0.89(3)  \\
1 & $5.3(4)\ten{-4}$ & $2.88(3)$ & $0.4$ & $0.629(12)$ & 3.27(15) \\
\hline \hline
\end{tabular}
\caption{Properties of the pion at zero temperature.
The index $n$ denotes the momentum ${\bf p}_n$ induced 
and $\omega^0_{{\bf p}_n}$ corresponds to the energy of the state (in particular, $\omega_{\bf 0}^0$ is the pion mass).
All errors are purely statistical, and all renormalization factors are included. 
The fit interval begins at $x_0/a=6$ for the zero-momentum case and at $x_0/a=15$ for one unit 
of momentum in view of the effective mass plot of Fig.~\ref{fig:u_T0}. 
Dimensionful quantities are normalized with $T=1/24a$.}
\label{tab:cosh_T0}
\end{table}

\subsection{The screening quantities $f_\pi$ and $m_\pi$ at finite temperature}
A detailed description of how the extraction is carried out can be
found in \cite{Brandt:2014qqa}. Here, we highlight the
basic relations that lead to the extraction of the screening
quantities $f_\pi$ and $m_\pi$ and therefore to the
values of the estimators $u_f$ and $u_m$.

\begin{figure}[h!]
\begin{center}
\includegraphics[width=.7\textwidth]{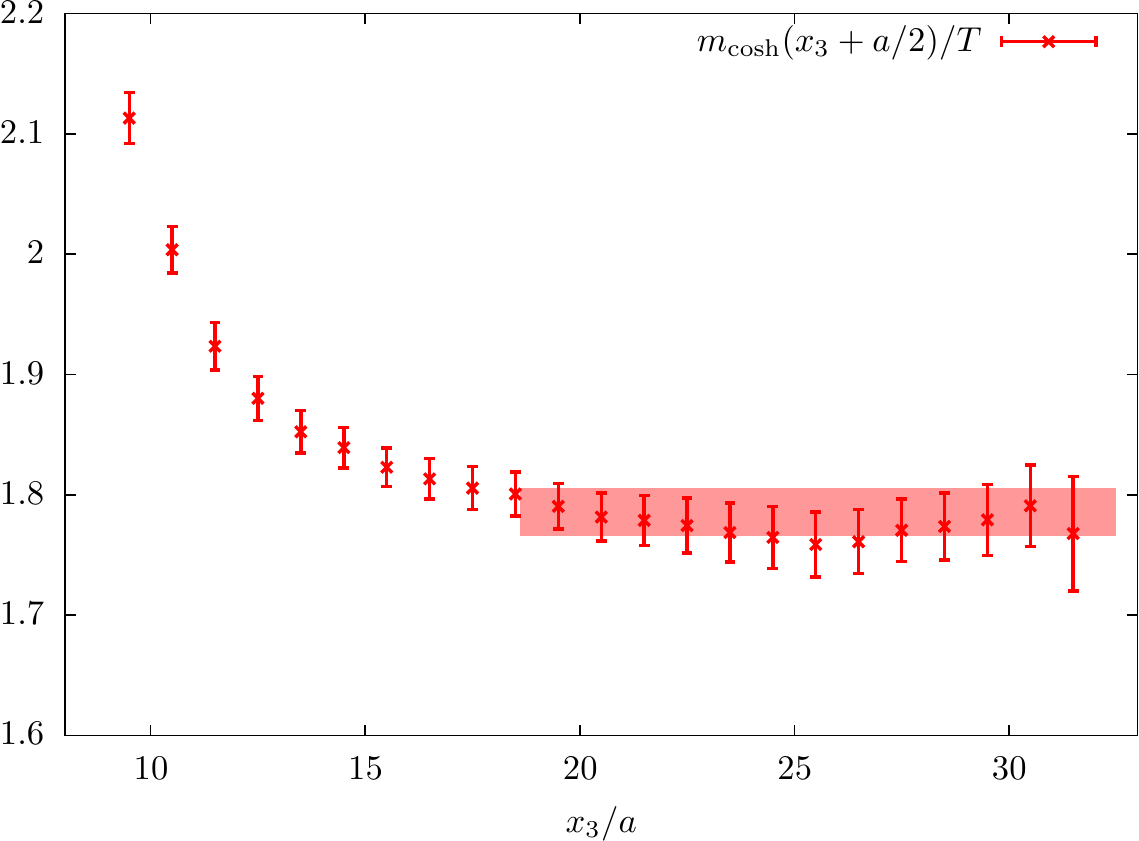}
\end{center}
\vspace{-0.3cm}
\caption{Effective `cosh' mass plot for the screening mass $m_\pi$. The plateau has been chosen to begin at $x_3/a=19$.}
\label{fig:coshmass_mpi}
\end{figure}

\begin{itemize}
\item The screening mass $m_\pi$ is calculated by fitting the
correlation function $G^{\text{s}}_P(x_3,T)$ with a two-state
ansatz of the form $ \sum_{i=1}^2 A_i\cosh[m_i (L/2-x_3)]$) with masses $m_i$
and amplitudes $A_i$.
The value obtained for the ground-state mass is compatible with the value obtained 
from the `cosh' mass which is defined as the positive root of the following equation (see Fig.~\ref{fig:coshmass_mpi}),
\be
\frac{G^{\text s}_P(x_3,T)}{G^{\text s}_P(x_3+a,T)} 
= \frac{\cosh[m_\text{cosh}(x_3+a/2)(x_3-L/2)]}{\cosh[m_\text{cosh}(x_3+a/2)(x_3+a-L/2)]}.
\ee
\item We determine the screening pion decay constant from the correlator $G^{\text{s}}_A(x_3,T)$ via \eqref{eq:Gsasy}
by applying again a two-state `cosh' ansatz.
The screening pion mass $m_\pi$ also appears in $G^{\text{s}}_A$.  
We use this fact as a consistency check, but due
to the better signal to noise ratio of the pseudoscalar channel, we
quote the value extracted from $G^\text{s}_P$ as our final result for $m_\pi$.
\end{itemize}

\subsection{Thermal time-dependent correlators at zero momentum \label{uf_um}}

The estimators $u_f$ and $u_m$ for the pion velocity $u$ are defined
in Eqs.\ (\ref{eq:um}-\ref{eq:uf}).  Apart from the PCAC mass, $f_\pi$
and $m_\pi$, they involve the time-dependent correlators
$G_A(x_0,T,{\bf 0})$ and $G_P(x_0,T,{\bf 0})$. The difference between
$u_f$ and $u_m$ can be explained as follows.  The estimator $u_m$ is
based on the dominance of the pion contribution in the correlators
$G_A$ and $G_P$ at $x_0=\beta/2$; the estimator $u_f$ is based on
assuming that the residue is given by the screening quantities (as
predicted by the chiral effective theory), ${\rm
  Res}(\omega_{\bf 0})=f_\pi^2 m_\pi^2$, and the dominance of the pion
contribution in $G_A$ only.  The dominance in $G_A$ is less strong an
assumption than the assumption that the pion dominates $G_P$, since
their spectral functions are related by
$\rho_P(\omega)=-\frac{\omega^2}{4m^2} \rho_A(\omega)$ at zero spatial
momentum.  We summarize results for $u_f$ and $u_m$ in
Table~\ref{tab:screening}.

\begin{figure}[t]
\begin{center}
\includegraphics[width=.7\textwidth]{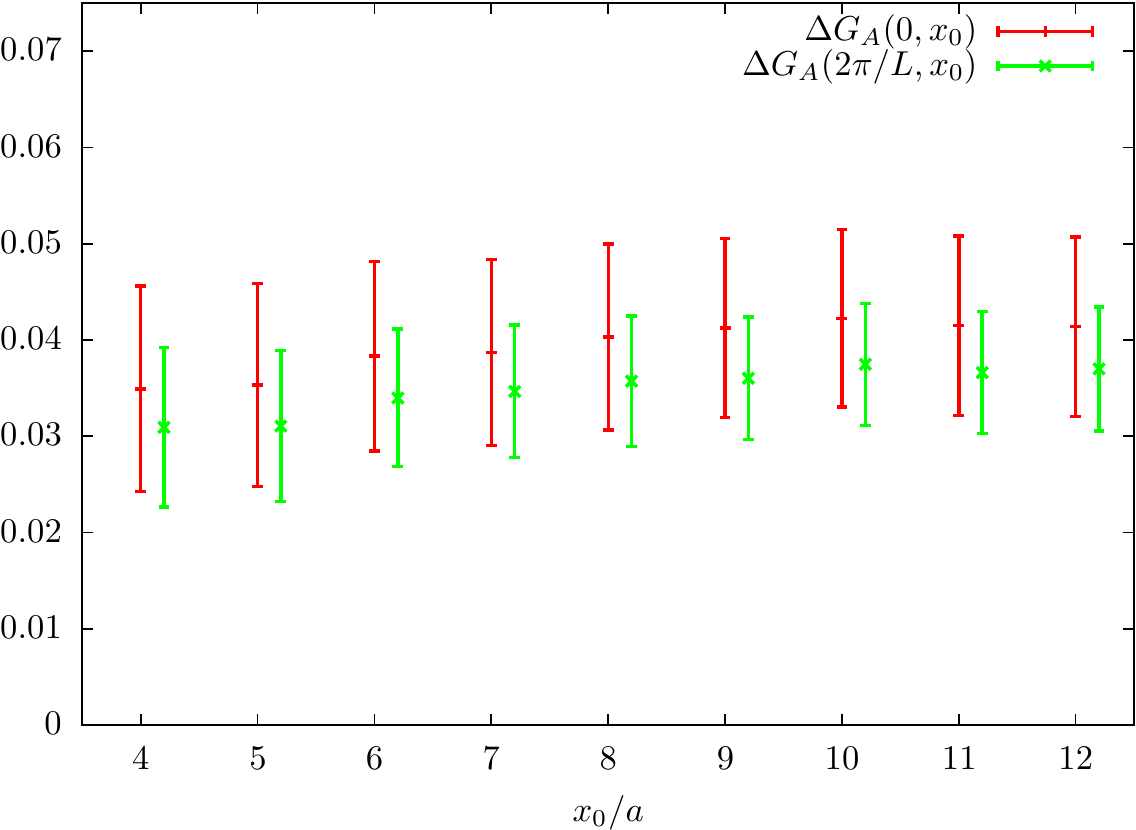}
\end{center}
\vspace{-0.3cm}
\caption{The difference $\Delta G_A(|{\bf p}|,x_0)\equiv [G_A(x_0,T,{\bf p}) - G_A^{\rm rec}(x_0,T,{\bf p})]/T^3$ of the 
  thermal correlator and the reconstructed correlator for $|{\bf p}|=0$ and $2\pi/L$.}
\label{fig:gsub}
\end{figure}

\begin{table}[h!]
\centering
\begin{tabular}{l@{~~~}r}
\hline \hline
$m_\pi/T$ & 1.79(2) \\ 
$f_\pi/T$ & 0.46(1) \\ \hline 
$u_f$ & 0.76(1) \\
$u_m$ & 0.74(1) \\
$u_f/u_m$ & 1.02(1) \\ \hline
$\omega_{\bf 0}/T$ & 1.32(2)\\
$f_{\pi}^t/T$ & 0.62(1) \\
$\text{Res}(\omega_{\bf 0})/T^4$ & 0.68(2)\\ \hline \hline
\end{tabular}
\caption{Summary of the results for the $N_\tau=24$ thermal
  ensemble. All
  renormalization factors are included and the errors are purely
  statistical. The value of $\omega_{\bf 0}$ is calculated using $\omega_{\bf 0}=u_m m_\pi$.
  In the same way $f_{\pi}^t=f_\pi/u_m$. The value of the
  residue is obtained according to Eq.\ (\ref{eq:res}),
  $\text{Res}(\omega_{\bf 0}) = f^2_\pi m^2_\pi$.}
\label{tab:screening}
\end{table}

The chiral expansion around $(T, m=0)$ proposed in \cite{Son:2002ci}
assumes that one is sufficiently close to the chiral limit. In this
limit, the screening pion mass $m_\pi$ vanishes and the coefficient $u(T)$ is indeed the velocity
of a massless pion quasiparticle in the presence of a thermal bath. A
deviation from unity corresponds to a violation of boost
invariance. At finite but small quark mass, we showed in~\cite{Brandt:2014qqa} that the 
consistency of $u_f$ and $u_m$ serves as an indicator for the applicability of the chiral
expansion. Based on the results of \tabref{tab:screening},
we conclude that they are indeed consistent.
The results, in particular for $u_f$, are in good agreement with the
results obtained in~\cite{Brandt:2014qqa} on ensembles with a coarser
lattice spacing and a slightly heavier quark mass.  Note that to
leading order, $u$ is expected to be independent of the quark mass.

The `reconstructed' correlator $G_A^{\rm rec}$ is defined as the
thermal Euclidean correlator that would be realized if the spectral
function remained the zero-temperature one. We compute it following
the method first proposed in~\cite{Meyer:2010ii}. Figure
\ref{fig:gsub} shows the difference between the thermal correlator and
the reconstructed correlator. There is a statistically significant
difference between the two correlators, which shows that a change must
take place in the spectral function. Because the difference is very
weakly dependent on time, the change must take place in the
low-frequency part of the spectral function.  We expect from the
thermal chiral effective theory that the change is due to a
modification of the mass and/or the residue of the pion quasiparticle.
Using the numbers of Table \ref{tab:screening}, the changes amount
respectively to
\be
\frac{\omega_{\bf 0}}{\omega_{\bf 0}^0} = 0.836(14),\qquad \quad 
\frac{f_{\pi}^t }{{f^0_\pi}}=  1.03(2).
\ee
We thus observe that while the pion decay constant remains unchanged
at the precision level of a few percent, the pion mass decreases by
$16\%$.  Qualitatively, these results are consistent with the two-loop
results in zero-temperature chiral perturbation theory given
in~\cite{Schenk:1993ru,Toublan:1997rr}. Future lattice calculations approaching the
chiral limit would allow for a quantitative comparison.

\subsection{The spectral function $\rho^{A}(\omega, p)$ at non-zero momentum}

As the next step, we test the functional form of
\eqref{eq:pion_disprel} at non-zero momentum. The relevant real-time
pion states with energy $\omega_{\bf p}$ have a non-zero overlap with
the operator $\int d^3x~ e^{i{\bf p x}} A_0(x)$; furthermore, the
spectral function $\rho_A$ becomes independent of $\omega$ in the
ultraviolet, rather than growing like $\omega^2$.  We therefore expect
to have the best sensitivity to the pion contribution in the
correlator $G_A(x_0, T, {\bf p})$.

At finite temperature, the analysis of the correlator $G_A(x_0,T,{\bf
  p})$ is more involved than at zero temperature: only at sufficiently
small quark masses and momenta, and not too small $x_0$ is the
correlator parametrically dominated by the pion pole. Therefore we
proceed by formulating a fit ansatz to take into account the non-pion
contributions.  The combination of Eqs.\ (\ref{eq:spec_T0}) and
(\ref{eq:asymptotic}) motivates an ansatz for the spectral function
reading
\be
\rho^A(\omega, {\bf p})  = A_1({\bf p})\sinh(\omega\beta/2) \delta(\omega-\omega_{\bf p}) + A_2({\bf p}) \frac{N_c}{24\pi^2}\left(1-e^{-\omega \beta}\right) \theta(\omega-c).
\ee 
The corresponding form of the correlation function then reads
\be
G^A(x_0, T, {\bf p}) =  A_1({\bf p}) \cosh(\omega_{\bf p} (\beta/2-x_0)) + A_2({\bf p}) \frac{N_c}{24\pi^2} \left(\frac{e^{-cx_0}}{x_0}+\frac{e^{-c(\beta-x_0)}}{\beta-x_0} \right).
\label{eq:fit_ansatz}
\ee

We fit $G_A(x_0, T, {\bf p})$ with the ansatz given in
\eqref{eq:fit_ansatz} for the momenta ${\bf p}_n = (0,0,2\pi n/L)$ with
$n=1,2,3,4,5$. The fit interval is chosen to be $x_0/a \in [5,12]$ in
order to avoid cutoff effects. There are four parameters involved,
$A_1({\bf p}),~ \omega_{\bf p},~ A_2({\bf p})$ and $c$.
Leaving $\omega_{\bf p}$ as a fit parameter led to poorly constrained fits.
 Therefore we set the value of $\omega_{\bf p}$ to the
prediction of \eqref{eq:pion_disprel} in order to test whether the data can be 
described in this way. Motivated by the expected large-$\omega$ behavior of the spectral function,
we quote the rescaled  parameter $\tilde{A_2} = A_2/{\bf p}^2$. Note that the  quark mass is negligible compared 
to all the non-vanishing $|{\bf p}|$ values considered here. 
The expected value of $\tilde{A_2}$ is of order unity, in view of \eqref{eq:asymptotic}.
Eq.\ (\ref{eq:residue}) allows us to establish the relation between the fit
 parameter $A_1({\bf p})$ and the residue itself,
\begin{eqnarray}
\text{Res}(\omega_{\bf p}) &=& 2A_1({\bf p})\omega_{\bf p}\sinh(\omega_{\bf p} \beta/2).
\end{eqnarray}
One may further convert the result for the residue into a parameter $b({\bf p})$,
defined by
\be
\text{Res}(\omega_{\bf p}) = f_\pi^2 (m_\pi^2 + {\bf p}^2) (1+b({\bf p})).
\ee
From the chiral prediction \eqref{eq:res},
we thus expect  $b({\bf p})$ to be small compared to unity if the effective description is working.
The results are summarized in \tabref{tab:u_m}.

\begin{table}[ht]
\centering
\begin{tabular}{c|c|c|c|c|c|c|c}
$n$ & $A_1/T^3$ & $\omega_{{\bf p}_n}/T$ & $\tilde{A_2}$ 
 & $c/T$ & $\text{Res}(\omega_{{\bf p}_n})/T^4$ & $b$ & $\chi^2/{\rm d.o.f}$ \\ \hline 
1 & $2.95(4)\ten{-1}$ & $2.19(3)$ & $1.78(8)$ & $6.7(3)$ & $1.72(6)$ & $-0.08(3)$ & $0.06$ \\
2 & $1.40(5)\ten{-1}$ & $3.73(6)$ & $1.26(2)$ & $6.1(1)$ &$3.3(2)$ & $-0.39(4)$ & $0.15$ \\
3 & $4.9(3)\ten{-2}$ & $5.40(9)$ & $1.19(1)$ & $7.7(1)$ & $3.9(5)$ & $-0.65(4)$ & $0.35$ \\
4 & $1.7(2)\ten{-2}$ & $7.1(1)$ & $1.15(1)$ & $9.67(9)$ & $4.21(7)$ &$-0.78(3)$& $0.49$ \\
5 & $4(1)\ten{-3}$ & $8.8(1)$ & $1.12(1)$ & $11.7(1)$ &$3(1)$ &$ -0.89(3) $& $1.04$ \\
\hline \hline
\end{tabular}
\caption{Results of fits to the axial-charge density correlator at non-vanishing momentum ${\bf p}_n$.
All errors quoted are statistical, and
all renormalization factors are included. The quantity $\omega_{\bf p}/T$ is not a fit parameter;
rather it is set to the value predicted by \eqref{eq:pion_disprel} with $u(T)=u_m=0.74(1)$.}
\label{tab:u_m}
\end{table}

\begin{figure}[h!]
\begin{center}
\includegraphics[width=.55\textwidth]{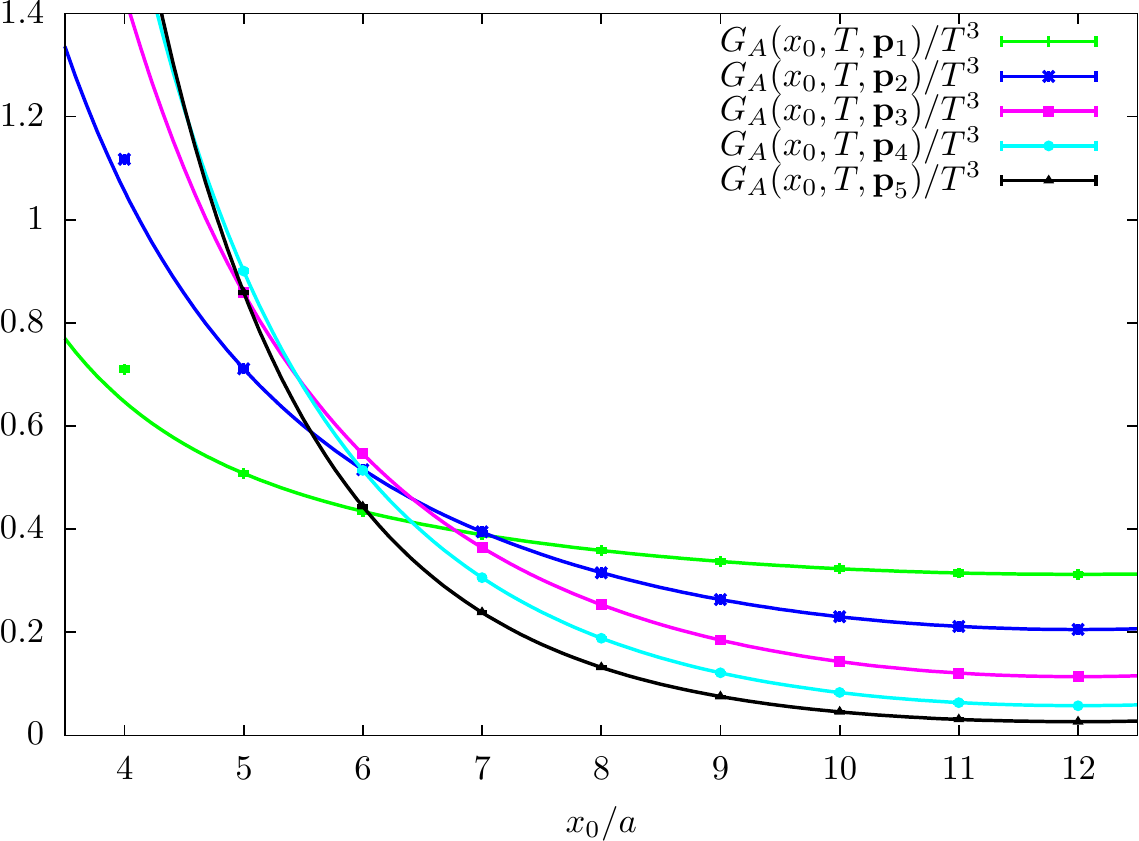}
\end{center}
\caption{Correlation functions $G_A(x_0,T, {\bf p}_n)/T^3$ 
with ${\bf p}_n=(0,0,2\pi n/L)$, together with the fits resulting from the four-parameter
Ansatz of \eqref{eq:fit_ansatz}. The corresponding parameter values are given in \tabref{tab:u_m}. 
All renormalization constants are included.}  
\label{fig:corr:finitek}
\end{figure}

The fits provide a good description of the data; see the $\chi^2/{\rm d.o.f}$  values and
Fig.\ \ref{fig:corr:finitek}.  We observe that at the smallest
momentum, $|{\bf p}|\simeq 400\,{\rm MeV}$, $b({\bf p})$ really is
small, pointing to a successful check of the chiral prediction.  At
higher momenta, the negative, order unity value of $b({\bf p})$ indicates that the
residue of the pion pole is reduced. It should also be remembered that
at higher momenta, neglecting the width of the quasiparticle is bound
to be an increasingly poor approximation.
The coefficient $\tilde{A_2}$ is expected to be of order unity from
the treelevel prediction \eqref{eq:asymptotic}. Indeed the numbers in \tabref{tab:u_m}
are of order unity. One reason for the relatively large value of 
the coefficient at the smallest momentum could be that axial-vector excitations
are contributing around the threshold $c$, thus adding spectral weight.
The value of the threshold at $|{\bf p}|\simeq 400\,{\rm MeV}$, is about 1.1GeV, 
a value we consider to be reasonable given that the mass of the lightest axial-vector 
meson in nature is $m_{a_1} \approx 1.2{\rm GeV}$.

In order to gauge the discriminative power of the test, it is
interesting to ask whether a rather different model is consistent with
the lattice data on $G_A(x_0,{\bf p},T)$.  We assume for this purpose
that the dispersion relation and the residue have the same ${\bf
  p}$-dependence as at zero temperature. We therefore set $\omega_{\bf
  p} = \sqrt{\omega_{\bf 0}^2 + {\bf p^2}}$, and obtain for $n=1$ an
equally good description of the data, with a value of the residue
${\rm Res}(\omega_{\bf p}) = 3.01(4)$ not too different from
$(f_{\pi}^t)^2 (\omega_{\bf 0}^2 + {\bf p^2})=2.84(7)$.  The other fit
parameters take the values $\tilde{A_2}= 2.42(17)$ and $c/T =
10.2(4)$.  While the perturbative coefficient and the threshold values
seem less plausible to us, we cannot completely exclude this model on
the basis of the lattice data.

To summarize, we have found that the dispersion relation of the pion
quasiparticle is consistent with \eqref{eq:pion_disprel}, the
parameter $u$ being determined at vanishing spatial momentum.  In
order to test the dependence of our results on the fit ansatz made, in
the next section we apply the Backus-Gilbert method.

\subsection{The Backus-Gilbert method for $\rho^A(\omega,{\bf p})$}

The Backus-Gilbert method is a method suitable for inverting integral
equations like \eqref{eq:spectral_def}. It has been studied in many
contexts (see e.g. \cite{Backus1,Backus2,Press:2007zz,
  doi:10.1080/01630568708816267,schomburg1987convergence,kirsch1988backus}). While
it has not been applied in lattice QCD, to our knowledge, the central
notion of resolution function was used in~\cite{Meyer:2007dy}.  We
first describe the method in some generality.  It is a completely
model-independent approach since no ansatz needs to be made for the
spectral function.

The goal is to solve the integral equation 
\begin{equation}
G(x_i) = \int^{\infty}_0 d\omega f(\omega) K(x_i,\omega) \qquad x_i \neq 0 \qquad \forall i
\end{equation}
for the unknown function $f(\omega)$, given the kernel $K(x_i,\omega)$ and given data on $G(x_i)$.
The idea is to define an estimator $\hat{f}(\bar\omega)$
\begin{equation}
\hat{f}(\bar\omega) = \int^\infty_0 \hat{\delta}(\bar\omega,\omega) f(\omega) d\omega
\label{eq:rho_est}
\end{equation}
where $\hat{\delta}(\bar\omega,\omega)$ is called the resolution function or
 averaging kernel. It is a smooth function concentrated around some reference value $\bar\omega$,
normalized according to $\int_0^\infty d\omega \hat\delta(\bar\omega,\omega) = 1$, and parametrized at fixed $\bar\omega$
by coefficients $q_i(\bar\omega)$,
\be
\hat\delta(\bar\omega,\omega) = \sum_i q_i(\bar\omega) K_i(\omega),
\ee
so that  $\hat f$ is obtained according to
\be
\hat{f}(\bar\omega) = \sum^n_{i=1} G(x_i) q_i(\bar\omega).
\label{eq:master_eq}
\ee
The goal is then to minimize the width of the resolution function. 
Minimizing the second moment of $\hat\delta(\bar\omega,\omega)$ in its second argument
around its first argument yields
\begin{equation}
q_i(\bar\omega) = \frac{\sum_j W^{-1}_{ij}(\bar\omega) R(x_j)}{\sum_{k,l} R(t_k)W^{-1}_{kl}(\bar\omega) R(x_l)},
\end{equation} 
where 
\begin{eqnarray}
W_{ij}(\bar\omega) &=& \int^\infty_0 d\omega K(x_i,\omega) (\omega-\bar\omega)^2 K(x_j, \omega),\\
R(x_i) &=& \int^\infty_0 K(x_i,\omega) d\omega.
\end{eqnarray}
We remark that $\hat{f}(\omega)$ equals $f(\omega)$ if the latter is constant.

The matrix $W_{ij}(\bar\omega)$ is very close to being
singular. This is the reason why, when trying to use a data set with
error bars, one needs to regulate and change the matrix $W$ to
\begin{equation}\la{eq:reg}
W_{ij} \to \lambda W_{ij} + (1-\lambda) S_{ij}, \qquad 0 < \lambda < 1
\end{equation}
where $S_{ij}$ is the covariance matrix of the data. The value of
$\lambda$ controls the trade-off between resolution and stability. For
values of $\lambda$ close to 1, we obtain the best possible resolution.
However the results tend to be  unstable since the matrix is poorly conditioned 
and large cancellations take place among the terms in \eqref{eq:master_eq}.
Reducing $\lambda$ improves the stability of the result at the cost 
of deteriorating the frequency resolution.

\begin{figure}[ht]
\begin{center}
\includegraphics[width=0.48\textwidth]{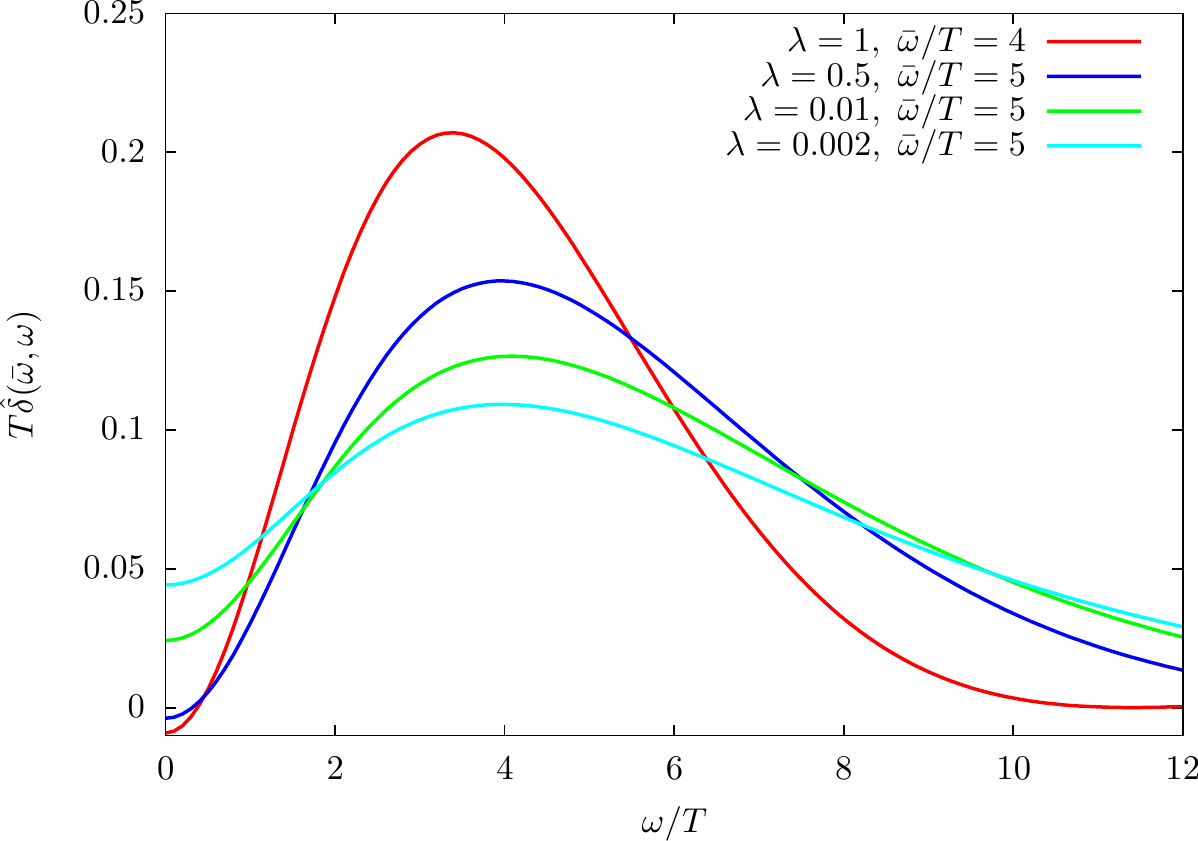}
\includegraphics[width=0.48\textwidth]{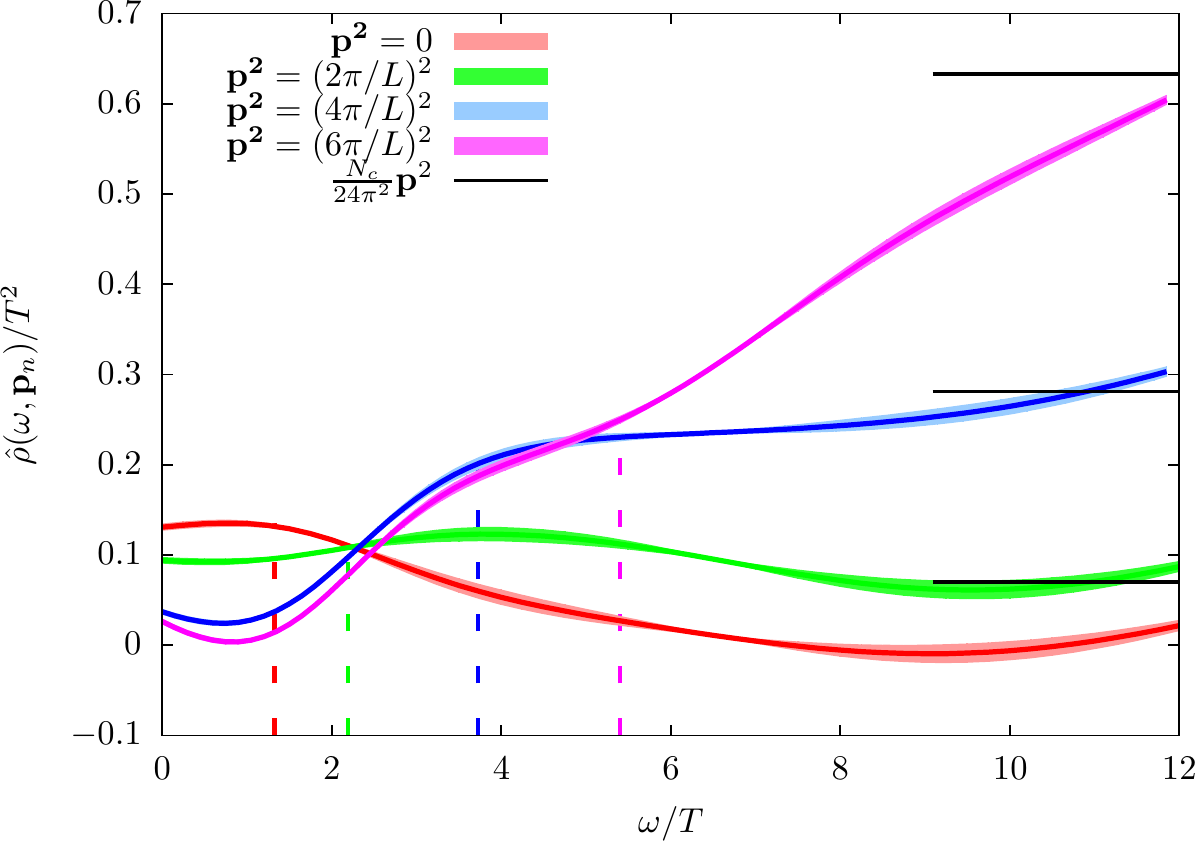}
\end{center} 
\caption{Left: Some examples of resolution functions for
  different values of $\lambda$ centered at $\bar\omega/T$. Right:
  Estimators $\hat{\rho}(\omega, {\bf p}_n)/T^2$ for $n=0,1,2,3$
  together with its error shown as a band. The vertical colored dashed
  lines correspond to the locations of the expected positions of the
  poles $\omega_{{\bf p}_n}$ according to \eqref{eq:pion_disprel} with
  $u(T) = u_m$. The black horizontal lines correspond to the
  treelevel asymptotic values of $\rho^A(\omega, {\bf p})$. All renormalization
  constants have been taken into account as well as the improvement
  program on the axial correlators. Dimensionful quantities
  have been made dimensionless by the appropriate power of $T=1/24a$.}
\label{fig:BG}
\end{figure}

We apply this method to \eqref{eq:spectral_def}. In order to regularize the finite-temperature 
kernel at $\omega=0$ we rewrite the equation as
\be
G_A(x_0, T, {\bf p}) = \int^\infty_0 d\omega\left(\frac{\rho^{A}(\omega,{\bf p})}{\tanh(\omega/2)}\right) \underbrace{\left(\frac{\cosh(\omega(\beta/2-x_0))}{\cosh(\omega \beta/2)}\right)}_{\doteq K(x_0,\omega)}.
\label{eq:mod_kernel_finiteT}
\ee
This defines our estimator $\hat{\rho}$ at $\omega = \bar\omega$,
\be
\hat{\rho}(\bar\omega, {\bf p}) = \int^\infty_0 d\omega \;\hat{\delta}(\bar\omega, \omega)\left( \frac{\rho_A(\omega,{\bf p})}{\tanh(\omega \beta/2)}\right).
\label{eq:hat_rho}
\ee
After  regulating the problem via the covariance matrix
$S_{ij}$ as in \eqref{eq:reg}, the inversion is carried out via Singular Value
Decomposition. This offers the opportunity to diagnose how badly
conditioned the matrix is. With all quantities
made dimensionless by applying appropriate powers of the temperature,
we choose $\lambda= 2\cdot 10^{-3}$ in the following.
Typical condition numbers of the regularized matrix in \eqref{eq:reg} are $\sim 10^{8}$. The situation gets worse when
$\lambda$ approaches unity, as explained above.  The results for zero
momentum and the first three units of momentum are shown in the right
panel of Fig.~\ref{fig:BG}. As in the case of the fit, we included the
points of the correlation in the interval $x_0/a=[5,12]$ so
$W_{ij}(\bar\omega)$ is a $n\times n$ symmetric matrix with $n=8$. 
 The
value of $\lambda$ was chosen equal to 0.002. This yields a relative
error on $\hat{\rho}$ of $\sim 3-5$\%, while the resolution function is displayed 
in the left panel of Fig.~\ref{fig:BG}. One direct observation is the
fact that the expected asymptotic behavior for large values of $\omega$ is
reproduced very well. 

\begin{figure}[h]
\begin{center}
\includegraphics[width=0.48\textwidth]{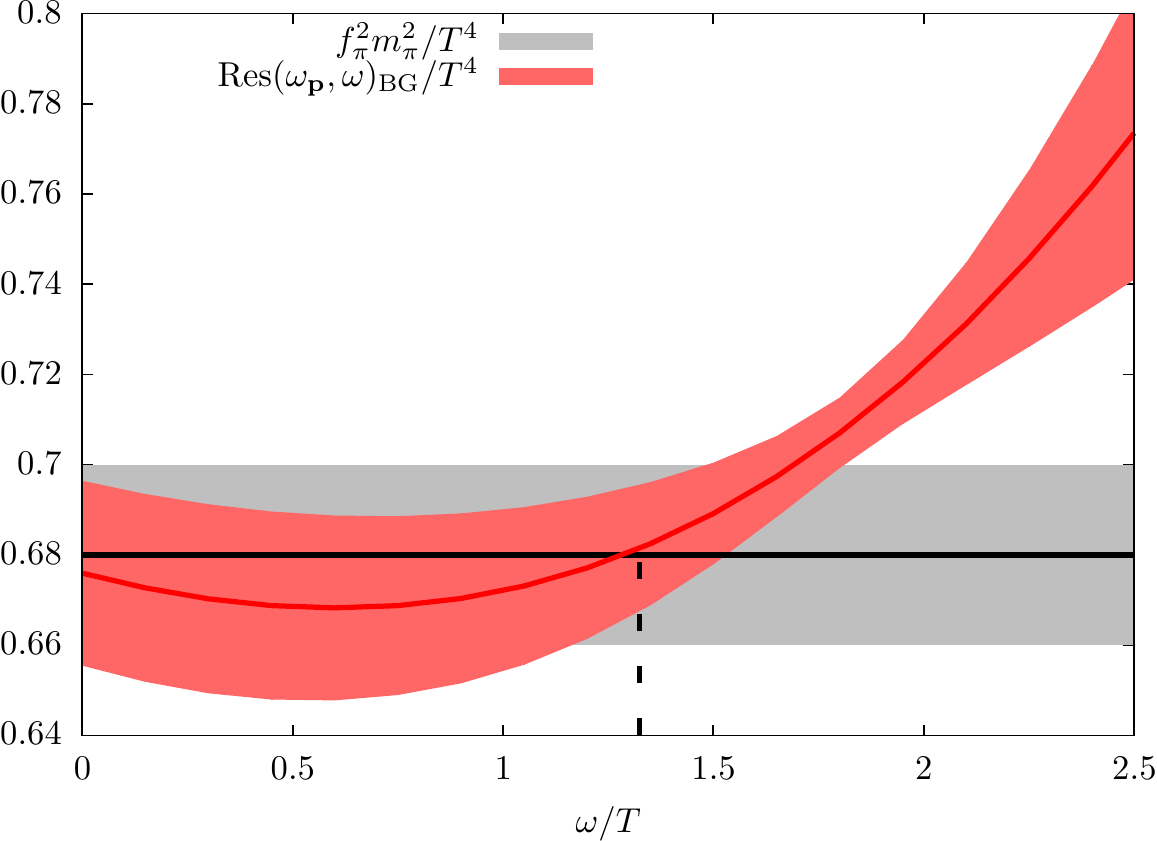}
\includegraphics[width=0.48\textwidth]{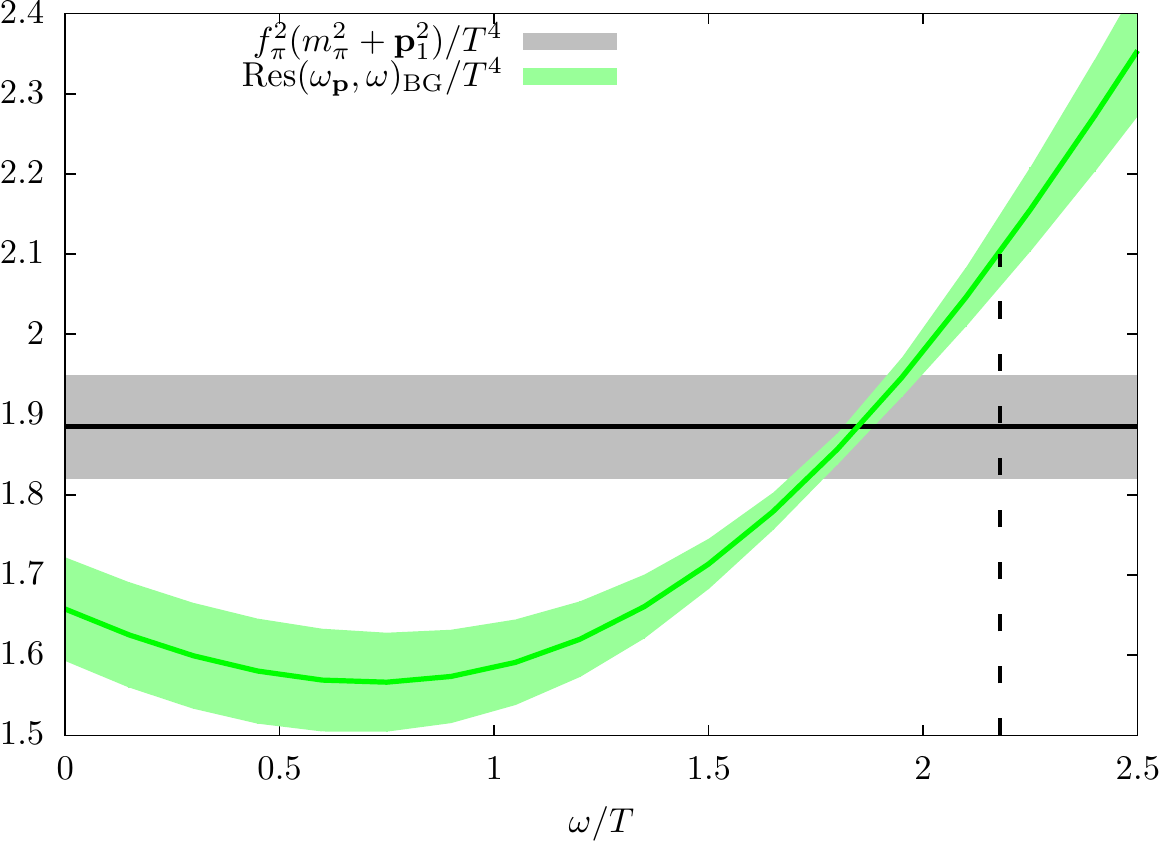}
\end{center}
\caption{The effective residue $\text{Res}(\omega_{\bf p}, \omega)_\text{BG}$ as
  defined in \eqref{eq:res_BG}. Left: No momentum induced, ${\bf
    p}=0$. Right: One unit of momentum induced, ${\bf p}_1 =
  (0,0,2\pi/L)$. The grey band is the expectation in terms of
  screening quantities, \eqref{eq:res}. All renormalization factors are
  included. The errors arise from the statistical uncertainty. The values 
of $\omega_{\bf p}$ are indicated by dashed vertical lines.}
\label{fig:res}
\end{figure}

The right panel of Fig.\ \ref{fig:BG} also shows the expected
positions of the poles that follow from \eqref{eq:pion_disprel} as
vertical colored dashed lines.  We now want to test the ${\bf p}$
dependence of the residue $\text{Res}(\omega_{\bf p})$ via the
following argument.  If we assume that, for a given value of $\omega$,
$\hat\rho(\omega,{\bf p})$ is dominated by the pion pole, we obtain
the following estimator for the residue,
\be
\text{Res}(\omega_{\bf p}, \omega)_\text{BG}= \frac{2\omega_{\bf p}
 \tanh(\omega_{\bf p} \beta/2)\hat{\rho}(\omega, {\bf p})}{\hat{\delta}(\omega,\omega_{\bf p})}.
\label{eq:res_BG}
\ee
Here we treat $\omega_{\bf p}$ as input and calculate it using
\eqref{eq:pion_disprel} with the value of $u=u_m=0.74(1)$ determined
at zero momentum. The result as a function of $\omega$ is shown in
Fig.~\ref{fig:res} for zero and one unit of momentum. The natural
choice where $\text{Res}(\omega_{\bf p}, \omega)_\text{BG}$ is
expected to be the best estimator of the residue is at $\omega \approx
\omega_{\bf p}$. Looking at Fig.~\ref{fig:res}, one sees that
approximately around this value the curve intercepts the grey band,
which represents the prediction \eqref{eq:res}. The latter is
particularly well verified at zero momentum, while the agreement at
$|{\bf p}|=|{\bf p}_1|\approx 400{\rm MeV}$ is at the ten percent
level. These observations provide a further test that the pion
dispersion relation \eqref{eq:pion_disprel} predicted by the thermal
chiral effective theory is consistent with the lattice two-point
function of the axial charge density.

Comparing the method followed in this subsection with the previous
method based on a global fit to the spectral function, the former has
the advantage of not requiring an explicit parametrization of the
non-pion contributions to the spectral function. This observation may
be useful in other lattice studies of spectral functions.

\section{Conclusion\label{sec:concl}}

We have found that the pion quasiparticle mass is reduced significantly by thermal
effects compared to its vacuum value -- unlike the pion screening
mass, which increases.  
Also, the energy cost of giving the pion quasiparticle a
momentum is significantly reduced, since the `velocity' is well below 
unity, $u\approx 0.74$.
We have tested that the pion indeed admits a modified dispersion
relation, \eqref{eq:pion_disprel}, by analyzing lattice two-point
functions.  The test is based on requiring the consistency with the
lattice data of the combined chiral prediction for the dispersion
relation and the residue of the pion pole in the two-point function of
the axial-charge density.
These conclusions could be strengthened further by repeating the
calculation at smaller quark masses and with higher
statistics. Discretization errors should also be studied. Having a
higher resolution in momentum could help in assessing the region of
validity of the chiral effective theory.  

Assuming the results hold to further scrutiny, one may wonder how
much a modified mass and dispersion relation of the pion affects (a) the
freeze-out mechanism in heavy-ion collisions and (b) the predictions
of the hadron resonance gas model for equilibrium properties.  In
answering the latter question, one must take into account that the
modification of the pion dispersion relation is due to the presence of
hadrons in the medium, and issues of double counting arise. However,
the following estimates may provide a useful first idea of the size of
the effect. At the temperature of 169MeV in the two-flavor theory that
we have been discussing, with a zero-temperature pion mass of 270MeV,
we estimate, using the hadron resonance gas model, an isovector quark
number susceptibility\footnote{The current is here normalized as $\sqrt{2}V_\mu^a(x)$,
see \eqref{eq:currdef}.}
amounting to $\chi_s/T^2 = 0.42$.  In the HRG model, the pion
contributes\footnote{Here $f_B(p) = (e^{\beta\omega_p}-1)^{-1}$ is the
  Bose distribution.}  $\chi_s/T^2|_{\rm pions} = 4\beta^3\int
\frac{d^3p}{(2\pi)^3} f_B(p)(1+f_B(p)) =0.28$.  If the
spatial-momentum integral in the pion contribution is cut off at
$p_{\rm max} = 400\,{\rm MeV}$ (roughly the range of validity of the
chiral effective theory that we found), the contribution is reduced to
$0.11$. If we instead use the modified dispersion relation with the
lower quasiparticle mass $\omega_{\bf 0}=223{\rm MeV}$ and $u=0.74$, the
contribution for $p<p_{\rm max}$ amounts again to $0.28$. It is
unclear whether one should include a contribution from higher momenta,
given that the thermal width of the pion may then not be
negligible. The numbers above illustrate that the contribution of the
pion to the quark number susceptibility might not be as strongly
affected as one may at first think. However, the contribution comes
from softer pions, which implies a reduced amplitude of the transport
peak in the two-point function of the vector current $V_i^a(x)$, an
effect that can be tested in lattice simulations.

Determining the dispersion relation of a non-Goldstone hadron would be
interesting to see whether the relatively strong change we have seen
in the pion properties is specific to chiral dynamics. 
In general, a kinetic theory description allows one to use as primary
degrees of freedom the quasiparticles specific to the temperature of
interest.  It is therefore much broader in applicability than the
hadron resonance gas model, but requires input information on the
quasiparticles.
The channel treated here illustrates the importance of having guidance from an
effective theory in reconstructing the gross features of the spectral
function. In doing so, applying the Backus-Gilbert method in a first
step can be useful in narrowing down the region of frequency where a
specific ansatz for the spectral function must be made.

\acknowledgments
{ We are grateful for the access to the
  zero-temperature ensemble used here, made available to us through
  CLS, and for the zero-temperature correlation functions generated within the Mainz lattice group. 
  We acknowledge the use of computing
  time for the generation of the gauge configurations on the \emph{JUGENE} and \emph{JUQUEEN}
  computers of the Gauss Centre for Supercomputing located at
  Forschungszentrum J\"ulich, Germany, allocated through the John von Neumann Institute for Computing (NIC) within project HMZ21.
  Part of the configurations and all correlation functions were computed on the
  dedicated QCD platforms ``\emph{Wilson}'' at the Institute for Nuclear
  Physics, University of Mainz, and ``\emph{Clover}'' at the Helmholtz-Institut Mainz. This work was supported by the
  \emph{Center for Computational Sciences in Mainz} as part of the
  Rhineland-Palatinate Research Initiative and by the DFG grant ME
  3622/2-1 \emph{Static and dynamic properties of QCD at finite
    temperature}.
}

\appendix

\section{Tensor structure of the axial current two-point functions\label{sec:tensor}}

We work in the Euclidean field theory and define the correlation function in momentum space as
\be\la{eq:PolTens}
\delta^{ab}\, \Pi^{\rm A}_{\mu\nu}(\hat\epsilon,k) \equiv \int d^4x \, e^{ik\cdot x} \<A^a_\mu(x) A^b_\nu(0)\>_{\hat\epsilon}.
\ee
The unit vector $\hat\epsilon$ points in the direction that defines the thermal boundary condition. It is 
$\hat\epsilon=(1,0,0,0)$ in the rest frame of the thermal system. By doing the change of integration variables
$x\to -x$ and using translation invariance
\be
\<A_\mu(-x)A_\nu(0)\>_{\hat\epsilon} = \<A_\nu(x)A_\mu(0)\>_{\hat\epsilon} 
\ee
and symmetry under the O(4) rotation $x\to -x$, 
\be
\<A_\mu(-x)A_\nu(0)\>_{\hat\epsilon} = \<A_\mu(x)A_\nu(0)\>_{-\hat\epsilon},
\ee
we have the symmetries
\be\la{eq:sym}
\Pi^{\rm A}_{\mu\nu}(\hat\epsilon,k)= \Pi^{\rm A}_{\nu\mu}(\hat\epsilon,-k) = \Pi^{\rm A}_{\mu\nu}(-\hat\epsilon,-k).
\ee
To write down the tensor decomposition, we have the building blocks
$\delta_{\mu\nu}$, $k_\mu$ and $\hat\epsilon_\mu$ at our disposal.
We can write down four structures that respect the symmetries (\ref{eq:sym}),
\be
\delta_{\mu\nu},\qquad \frac{k_\mu k_\nu}{k^2},\qquad \frac{\hat\epsilon\cdot k}{k^2}(\hat\epsilon_\mu k_\nu + k_\mu \hat\epsilon_\nu),
\qquad \hat\epsilon_{\mu}\hat\epsilon_{\nu}.
\ee
We can form one  projector to the subspace orthogonal to both $\hat\epsilon$ and $k$, 
\be
C_{\mu\nu}^{\rm T,t} = \delta_{\mu\nu} - \frac{1}{1-(\hat\epsilon\cdot k)^2/k^2}\Big(
\hat\epsilon_{\mu}\hat\epsilon_{\nu} + \frac{k_\mu k_\nu}{k^2} 
- \frac{1}{k^2} (\hat\epsilon\cdot k)(\hat\epsilon_\mu k_\nu + k_\mu \hat\epsilon_\nu) \Big),
\ee
and one projector onto the component of $\hat\epsilon$ which is orthogonal to $k$,
\be
C_{\mu\nu}^{\rm T,l}  = \delta_{\mu\nu} - \frac{k_\mu k _\nu}{k^2} - C_{\mu\nu}^{\rm T,t}
 =\frac{1}{1-(\hat\epsilon\cdot k)^2/k^2}
\Big(\hat\epsilon_\mu - \frac{(\hat\epsilon\cdot k)k_\mu}{k^2} \Big)\Big(\hat\epsilon_\nu - \frac{(\hat\epsilon\cdot k)k_\nu}{k^2} \Big).
\ee
Two possible non-transverse combinations are 
\be
C_{\mu\nu}^{\rm L,l} =\frac{k_\mu k_\nu}{k^2}, 
\qquad 
C_{\mu\nu}^{\rm M} = \frac{1}{1-(\hat\epsilon\cdot k)^2/k^2}
\Big(\hat\epsilon_{\mu}\hat\epsilon_{\nu} -(\hat\epsilon\cdot k)^2 \frac{k_\mu k_\nu}{(k^2)^2}\Big).
\ee
The first one is the projector onto the direction of $k_\mu$.
The second tensor, while not a projector, has the properties
\be
C_{\mu\mu}^{\rm M} = 1,\qquad \quad 
C_{\mu\alpha}^{\rm T,t} \; C_{\alpha\nu}^{\rm M} = C_{\mu\alpha}^{\rm L,l}\; C_{\alpha\nu}^{\rm M}  = 0.
\ee
In summary, we can write
\be\la{eq:decomp}
\Pi^{\rm A}_{\mu\nu}(\hat\epsilon,k)= C_{\mu\nu}^{\rm T,t} \;\Pi^{\rm T,t} + C_{\mu\nu}^{\rm T,l} \;\Pi^{\rm T,l} 
+ C_{\mu\nu}^{\rm L,l} \;\Pi^{\rm L,l} + C_{\mu\nu}^{\rm M} \;\Pi^{\rm M}.
\ee
The argument of the $C$'s is $(\hat\epsilon,k)$, while the argument of the form factors $\Pi$ is 
$(\hat\epsilon\cdot k,k^2)$.

It is helpful to be able to invert the relation Eq.\ (\ref{eq:decomp}) in order to project out the 
form factors individually. We find
\ba
\la{eq:pLl}
\Pi^{\rm L,l}& =& \frac{k_\mu k_\nu}{k^2} \Pi_{\mu\nu}^A,
\\
\la{eq:pM}
\Pi^{\rm M}& =&  \frac{1}{\hat\epsilon\cdot k}\; k_\mu \Pi_{\mu\nu}^A 
(\hat\epsilon_\nu - \frac{\hat\epsilon\cdot k}{k^2}k_\nu),
\\
\la{eq:pTl}
\Pi^{\rm T,l}& =& 
\frac{1}{ 1-(\hat\epsilon\cdot k)^2/k^2}\;
\Big[  \hat\epsilon_\mu \Pi_{\mu\nu}^A\hat\epsilon_\nu - \frac{(\hat\epsilon\cdot k)^2}{(k^2)} \Pi^{\rm L,l}
- (1+ (\hat\epsilon\cdot k)^2/k^2) \,\Pi^{\rm M}\Big],
\\
\la{eq:pTt}
\Pi^{\rm T,t}& =& \frac{1}{2}\Big\{ \Pi_{\mu\mu}^A - 
\Big[  \Pi^{\rm T,l} +  \Pi^{\rm M} + \Pi^{\rm L,l}\Big]\Big\}.
\ea

\subsection{Special kinematics}

When $(\hat\epsilon\cdot k)^2 = k^2$, corresponding to vanishing spatial momentum 
in the rest frame of the thermal system, the projectors $C^{\rm T,t}_{\mu\nu}$ and  $C^{\rm T,l}_{\mu\nu}$
as well as $C^{\rm M}_{\mu\nu}$ become singular. Therefore we will define the value of the form factors in this limit  by continuity.
When $\hat\epsilon$ and $k$ are collinear, there are only two independent tensor structures,
\be
\Pi^{A,{\rm col}}_{\mu\nu}(\hat\epsilon,k) = \Big(\delta_{\mu\nu}  -  \frac{k_\mu k_\nu}{k^2}\Big) \hat\Pi_{\rm T}(k^2)
+ \frac{k_\mu k_\nu}{k^2} \hat\Pi_{\rm L}(k^2).
\ee
Applying the relevant projectors as in Eqs.\ (\ref{eq:pLl}--\ref{eq:pTt}), one finds that
\ba
\Pi^{\rm T,t} = \Pi^{\rm T,l} = \hat\Pi_{\rm T}, \qquad \quad \Pi^{\rm L,l}=\hat\Pi_{\rm L}, \qquad \quad \Pi^{\rm M}=0. 
\ea

When $\hat\epsilon\cdot k=0$, corresponding to the static correlators, $C^{\rm T,l}_{\mu\nu}$ becomes equal to $C^{\rm M}_{\mu\nu}$.
Therefore, in that situation the Euclidean correlator is only sensitive to the sum of the two corresponding
form factors, $(\Pi^{\rm M}+\Pi^{T,l})$. \eqref{eq:pM} nonetheless provides an unambiguous definition 
of $\Pi^{\rm M}$ if, expressed in the rest frame, $\lim_{k_0\to 0}  \Pi^A_{0i} / k_0$ is known.
The latter limit, however, requires an analytic continuation of the Euclidean correlator.

\subsection{The zero-temperature limit}

At zero temperature, it is natural to parametrize the correlation function as 
\be\la{eq:T0PiA}
\Pi^{\rm A}_{\mu\nu}(k)= \Big(\delta_{\mu\nu}  -  \frac{k_\mu k_\nu}{k^2}\Big) \Pi^{\rm T}(k^2)
+ \frac{k_\mu k_\nu}{k^2} \Pi^{\rm L}(k^2).
\ee
Applying the same projectors as in Eqs.\ (\ref{eq:pLl}--\ref{eq:pTt}) onto the correlation function (\ref{eq:T0PiA}),
and requiring that the same result be obtained in the zero-temperature limit, we obtain 
\ba
\Pi^{\rm L,l}(\hat\epsilon\cdot k,k^2) &\longrightarrow& \Pi^{\rm L}(k^2),
\\
\Pi^{\rm M}(\hat\epsilon\cdot k,k^2) &\longrightarrow& 0,
\\ 
\Pi^{\rm T,l}(\hat\epsilon\cdot k,k^2) &\longrightarrow& \Pi^{\rm T}(k^2),
\\
\Pi^{\rm T,t}(\hat\epsilon\cdot k,k^2) &\longrightarrow& \Pi^{\rm T}(k^2).
\ea

\subsection{Relation to the correlators of the pseudoscalar density}

We define
\ba
\delta^{ab}{\cal A}_\mu(\hat\epsilon,k) &=& \int d^4x\;e^{ikx}\; \<A^a_\mu(x)\;P^b(0)\>,
\\
\delta^{ab}{\cal P}(\hat\epsilon,k) &=& \int d^4x\;e^{ikx}\; \<P^a(x)\;P^b(0)\>.
\ea
We note the symmetry relations
\ba
{\cal A}_\mu(\hat\epsilon,k) = - {\cal A}_\mu(-\hat\epsilon,-k),
\\
\la{eq:AnuPtransl}
\int d^4x \; e^{ikx} \;\<P(x)\; A_\nu(0)\>_{\hat\epsilon} = {\cal A}_\nu(\hat\epsilon,-k),
\ea
respectively from O(4) invariance and from translation invariance.

In~\cite{Brandt:2014qqa}, Eq.\ (A7) and (A8), 
taking into account \eqref{eq:AnuPtransl} and \eqref{eq:sym},
the Ward identities
\ba
2m\, {\cal A}_\mu(\hat\epsilon,k) &=& ik_\alpha\;\Pi^{\rm A}_{\mu\alpha}(\hat\epsilon,k),
\\
4m^2\; {\cal P}(\hat\epsilon,k) &=& k_\mu \Pi^{\rm A}_{\mu\alpha}(\hat\epsilon,k) k_\alpha + m\<\bar\psi\psi\>
\ea
were derived. Inserting our tensor decomposition of $\Pi^{\rm A}_{\mu\alpha}(\hat\epsilon,k)$, we find 
\ba\la{eq:AmuRel}
2m\, {\cal A}_\mu(\hat\epsilon,k) &=& ik_\mu \;\Pi^{\rm L,l}(\hat\epsilon\cdot k,k^2) + i(\hat\epsilon\cdot k) 
\frac{\hat\epsilon_\mu - ({\hat\epsilon\cdot k}/{k^2})\; k_\mu}{1-(\hat\epsilon\cdot k)^2 /k^2} \; \Pi^{\rm M}(\hat\epsilon\cdot k,k^2),
\\  \la{eq:Prel}
4m^2\; {\cal P}(\hat\epsilon,k) &=& k^2\; \Pi^{\rm L,l}(\hat\epsilon\cdot k,k^2) + m\,\<\bar\psi\psi\>.
\ea

\section{On the residue of the pion pole\label{sec:apdx_residue}}

In this appendix, we use the general results of the previous section 
in the rest frame of the thermal system, $\hat\epsilon_\mu=(1,0,0,0)$.
The form factors are thus functions of $k_0$ and $k^2$ and the dependence 
on $\hat\epsilon$ is no longer indicated explicitly.
All expressions for correlation functions in this section refer exclusively
to the pion contribution.

In \cite{Brandt:2014qqa}, it was shown that 
the residue of the pion pole in the two-point function of $A_0$
at vanishing spatial momentum is ${\rm Res}(\omega_{\bf 0})=f_\pi^2 m_\pi^2$.
In order to determine the form of the residue at finite momentum, we 
parametrize the residue as 
\be
{\rm Res}(\omega_{\bf k})=f_\pi^2 (m_\pi^2 + \lambda {\bf k}^2).
\ee
To determine the parameter $\lambda$, we will exploit the fact that
the spectral representation of the two-point function of $A_0$ in
terms of real-time excitations must agree with the spectral
representation in terms of screening states.  From the former point of
view, the pion contribution to the correlator in momentum space takes
the form
\be\la{eq:Pi00t}
\Pi_{00}^A(k) = \frac{f_\pi^2 (m_\pi^2 + \lambda {\bf k}^2)}{k_0^2+\omega_{\bf k}^2},
\ee
with $\omega_{\bf k}$ given in \eqref{eq:pion_disprel}.
From the `screening' point of view, the residue must be proportional to $k_0^2$
at small $k_0^2$ (here we invoke the analytic continuation in the frequency, away
from the Matsubara values $k_0 = 2\pi T n$). This is so because the screening pion 
is odd under the Euclidean time reversal $x_0\to -x_0$, while $A_0$ is even. Thus we
can write
\be\la{eq:Pi00s}
\Pi_{00}^A(k) = \frac{ -|F|^2 k_0^2}{{\bf k}^2 + \frac{k_0^2}{u^2} + m_\pi^2}
\ee
for some parameter $F$ to be determined.
Equations (\ref{eq:Pi00t}) and (\ref{eq:Pi00s}) must agree when the numerators
are evaluated at the pole, $k_0^2 = - \omega_{\bf k}^2$. From here we learn the following,
\be
|F| = \frac{f_\pi}{u^2}, \qquad \qquad \lambda = 1.
\ee
This shows in particular that the residue has the form given in \eqref{eq:res}.
Essentially the same argument was already used in \cite{Brandt:2014qqa} to determine the residue of 
the pion pole in the two-point function of the pseudoscalar density,
\be
{\cal P}(k) =  - \frac{\<\bar\psi\psi\>^2 \,u^2}{4f_\pi^2} \,\frac{1}{k_0^2+\omega_{\bf k}^2}.
\ee
We note that for a one-pole contribution, factorization relations such as 
\be
|{\cal A}_0(k)|^2 = |{\cal P}(k)|\; |\Pi_{00}^A(k)|
\ee
hold. The phase of ${\cal A}_0$ can then be determined through its form at vanishing spatial momentum
given in \cite{Brandt:2014qqa}.

\subsection{The pion contribution to $\Pi_{\mu\nu}^A$}
 
Having found the residue of the pion pole in the various two-point functions of the axial current,
we give for completeness the pion contribution to the form factors defined in Eqs.\ (\ref{eq:pLl}--\ref{eq:pTt}),
\ba
\Pi^{\rm L,l}(k_0,{ k}^2) &=& -\frac{f_\pi^2 m_\pi^4 u^2}{k^2(k_0^2+\omega_{\bf k}^2)},
\\
\Pi^{\rm M}(k_0,{ k}^2) &=& \frac{f_\pi^2 m_\pi^2 {\bf k}^2(1-u^2)}{k^2(k_0^2+\omega_{\bf k}^2)},
\\
\Pi^{\rm T,l}(k_0,{ k}^2) &=& \frac{f_\pi^2{\bf k}^2(1-u^2)}{k_0^2+\omega_{\bf k}^2},
\\
\Pi^{\rm T,t}(k_0,{ k}^2) &=& 0.
\ea
The first is obtained from \eqref{eq:Prel}, then the second from
\eqref{eq:AmuRel}, the third by using \eqref{eq:pTl} and the first two
results.  Via \eqref{eq:decomp}, the form factors allow one to obtain
the entire tensor $\Pi_{\mu\nu}^A$.

These calculations could be greatly expedited by using an effective
Lagrangian, as written down in~\cite{Son:2002ci}. However it is also
instructive to derive the results above directly within QCD.


\section{Lattice correlation functions\la{sec:tables}}

\vspace{0.5cm}

\begin{table}[h!]
\centering
\begin{tabular}{c|c|c|c|c|c|c}
$x_0/a$ & $n=0$ & $n=1$ & $n=2$ & $n=3$ & $n=4$ & $n=5$ \\ \hline
0 & $2.9921(9)\ten{3}$ & $2.9962(9)\ten{3}$ & $3.0079(9)\ten{3}$ & $3.0260(9)\ten{3}$ & $3.0496(9)\ten{3}$ & $3.0774(9)\ten{3}$\\
1 & $9.200(5)\ten{1}$ & $9.335(5)\ten{1}$ & $9.698(5)\ten{1}$ & $1.0213(5)\ten{2}$ & $1.0813(5)\ten{2}$ & $1.1447(5)\ten{2}$\\
2 & $5.65(1)\ten{0}$ & $6.27(1)\ten{0}$ & $7.83(1)\ten{0}$ & $9.77(1)\ten{0}$ & $1.171(1)\ten{1}$ & $1.343(1)\ten{1}$\\
3 & $1.566(6)\ten{0}$ & $1.903(5)\ten{0}$ & $2.681(5)\ten{0}$ & $3.527(4)\ten{0}$ & $4.222(4)\ten{0}$ & $4.699(5)\ten{0}$\\
4 & $5.19(5)\ten{-1}$ & $7.10(4)\ten{-1}$ & $1.117(3)\ten{0}$ & $1.491(3)\ten{0}$ & $1.721(3)\ten{0}$ & $1.801(3)\ten{0}$\\
5 & $4.02(5)\ten{-1}$ & $5.08(4)\ten{-1}$ & $7.11(3)\ten{-1}$ & $8.58(2)\ten{-1}$ & $9.00(2)\ten{-1}$ & $8.59(2)\ten{-1}$\\
6 & $3.84(5)\ten{-1}$ & $4.34(4)\ten{-1}$ & $5.16(3)\ten{-1}$ & $5.46(2)\ten{-1}$ & $5.14(1)\ten{-1}$ & $4.44(1)\ten{-1}$\\
7 & $3.77(5)\ten{-1}$ & $3.90(4)\ten{-1}$ & $3.95(3)\ten{-1}$ & $3.64(2)\ten{-1}$ & $3.06(1)\ten{-1}$ & $2.386(7)\ten{-1}$\\
8 & $3.72(5)\ten{-1}$ & $3.59(4)\ten{-1}$ & $3.16(2)\ten{-1}$ & $2.54(1)\ten{-1}$ & $1.887(9)\ten{-1}$ & $1.319(6)\ten{-1}$\\
9 & $3.68(5)\ten{-1}$ & $3.37(4)\ten{-1}$ & $2.64(2)\ten{-1}$ & $1.85(1)\ten{-1}$ & $1.217(8)\ten{-1}$ & $7.55(5)\ten{-2}$\\
10 & $3.66(5)\ten{-1}$ & $3.23(4)\ten{-1}$ & $2.30(2)\ten{-1}$ & $1.44(1)\ten{-1}$ & $8.35(8)\ten{-2}$ & $4.59(5)\ten{-2}$\\
11 & $3.65(5)\ten{-1}$ & $3.15(4)\ten{-1}$ & $2.11(2)\ten{-1}$ & $1.21(1)\ten{-1}$ & $6.37(7)\ten{-2}$ & $3.11(5)\ten{-2}$\\
12 & $3.65(5)\ten{-1}$ & $3.13(4)\ten{-1}$ & $2.06(2)\ten{-1}$ & $1.14(1)\ten{-1}$ & $5.77(8)\ten{-2}$ & $2.67(6)\ten{-2}$\\
\hline \hline
\end{tabular}
\caption{Renormalized correlation function $G_A(x_0, T, {\bf p}_n)/T^3$. All errors quoted are statistical.}
\label{tab:data}
\end{table}

\bibliography{/Users/harvey/BIBLIO/viscobib}

\end{document}